\documentclass[journal,twoside]{IEEEtran}
\usepackage{float}
\usepackage{graphicx}
\usepackage{color}
\usepackage[square, comma, sort&compress, numbers]{natbib}
\usepackage{hyperref}
\usepackage{subfigure}
\usepackage{color}
\usepackage{amsmath}
\usepackage{bm}
\usepackage{flushend}
\usepackage{algorithm}
\usepackage{algorithmic}
\usepackage{multicol}
\usepackage{booktabs}
\usepackage{makecell}
\usepackage{colortbl}
\usepackage{array,booktabs}
\newcolumntype{P}[1]{>{\centering\arraybackslash}p{#1}}
\definecolor{mygray}{gray}{.9}
  \newlength{\halfpagewidth}
    \divide\halfpagewidth by 2

\newtheorem{Thm}{Theorem}

\newtheorem{Cor}{Corollary}
\newtheorem{Def}{Definition}

\newtheorem{Rem}{Remark}

\begin{document}

\title{Synthesizing a Clock Signal with Reactions---\\Part I: Duty Cycle Implementation Based on Gears}
\author{Chuan~Zhang,~\IEEEmembership{Member,~IEEE},
        Lulu~Ge,~\IEEEmembership{Student Member,~IEEE},
        Xiaohu~You,~\IEEEmembership{Fellow,~IEEE}\\
\thanks{Chuan Zhang, Lulu Ge, and Xiaohu You are all with National Mobile Communications Research Laboratory, Southeast University, Nanjing, China. Email: \{chzhang, luluge, xhyu\}@seu.edu.cn.}
\thanks{This paper was presented in part at IEEE Workshop on Signal Processing Systems (SiPS), 2015. Chuan Zhang and Lulu Ge contributed equally to this work. \emph{(Corresponding author: Chuan Zhang.)}}
}

\markboth{IEEE transactions on xxx,~2017}%
{Chuan Zhang \MakeLowercase{\textit{et al.}}: Synthesizing a Clock Signal with Reactions---Part I: Duty Cycle Implementation Based on Gears}
\maketitle

\begin{abstract}
Timing is of fundamental importance in biology and our life. Borrowing ideas from mechanism, we map our clock signals onto a gear system, in pursuit of better depiction of a clock signal implemented with chemical reaction networks (CRNs). On a chassis of gear theory, more quantitative descriptions are offered for our method. Inspired by gears, our work to synthesize a tunable clock signal could be divided into two parts. Part I, this paper, mainly focuses on the implementation of clock signals with three types of duty cycles, namely $1/2$, $1/N$ ($N > 2$), and $M/N$. Part II devotes itself in addressing frequency alteration issues of clock signals. \textcolor{black}{Guaranteed by existing literature, the experimental chassis can be taken care of by DNA strand displacement reactions, which lay a solid foundation for the physical implementation of nearly arbitrary CRNs.}
\end{abstract}

\begin{IEEEkeywords}
Clock signal, chemical reaction networks (CRNs), gear systems, duty cycle.
\end{IEEEkeywords}

\IEEEpeerreviewmaketitle

\section{Introduction}\label{sec:1}
\IEEEPARstart{H}{arnessing} engineering principles to design biological components opens a new era of synthetic biology \cite{synbio}. Indeed, since synthetic biology gained its more modern usage in 1974, a significant quantity of achievements have been initialized---synthetic DNA \cite{kosuri2014large,blight2000efficient,gibson2008complete,luo2000synthetic}, designed proteins \cite{broglia1998folding,broglia1999stability}, bio-information storage \cite{church2012next,eigen1966chemical,black1987biochemistry}, biosensors \cite{turner1987biosensors,sassolas2008dna,turner2000biosensors}, and even synthetic life \cite{ray1993evolutionary,langton1997artificial,malyshev2014semi}. To date, exciting and novel ideas show this nascent field's amazing vigorousness and endless potential. However, to coordinate tasks more properly, both biology and engineering call for clock signals of synthetic biology. Though systematic methods for designing an electrical clock signal can be used for both hardware and software, it becomes inapplicable in synthetic biology due to implementation differences.


Actually, a plenty of literature devote themselves in synthesizing biochemical time. In \cite{franco2011timing}, Franco used a synthetic biochemical oscillator proposed by Kim and Winfree \cite{kim2011synthetic}, which consists of two DNA transcription templates and two output RNA species. Power analysis was carried out to make this oscillator capable of driving a given load. Although it does not offer a method to synthesize a biochemical clock signal, it provides a reference to study the driving strength of a biochemical oscillator. 
In \cite{cardelli2009artificial}, a method to implement a biological oscillator with positive feedback transitions was proposed. Unfortunately, those aforementioned literature did not offer a systematic method to synthesize a biological clock signal with CRNs. Though our previous work \cite{ge2015formal} proposed a synthesis methodology for a clock signal regarding duty cycle, it lacks systematic modeling, fundamental analysis, and generalization on both generation and operation. By far, for a more vivid depiction and a better understanding, it is required to find a physical analogy and formal design methodology for synthesizing a clock signal.

Amazingly, from a common sense---time is invisible but could be heard with the sound of ``tick-tack'' from a mechanical watch in our daily life. Thus gear systems turn out to be a candidate to offer a general clock design methodology. Apart from this intuition, there are another three reasons to focus on gear systems. First, purely observing or studying the CRNs may be too abstract, and the design steps remain unrevealed. Second, the existence of a wide range of literature on gear systems \cite{litak2005dynamics,walha2009nonlinear,gear} makes the theory and implementation of gear systems quite mature. Third, various resemblances between a gear system and a clock tree enables such literature good references for bio-clock signal design. Specifically, both gear system and clock tree aim at efficient energy transmission in a designed regularity. Therefore, with a fundamental model of gear system, a clock tree in CRNs could be better understood. All those understandings offer us guidance and feasibility to generalize the methodology for further and wider applications. \textcolor{black}{Thus, our study is initially motivated to better depict our previous work in \cite{ge2015formal} based on a newly proposed gear model, and is dedicated itself to implement a tunable CRN clock signal in terms of duty cycle and frequency. Part I offers a systematic method to implement nearly arbitrary duty cycle on a chassis of gear systems. Part II contemplates to handle frequency alteration. This paper mainly covers the first part.}


The remainder of this paper is organized as follows. Section \ref{sec:2} briefly reviews the prerequisite preliminaries. Section \ref{sec:3} illustrates inspirations from a gear system to a CRN clock tree. Case studies are also given in this section. Section \ref{sec:4} presents the design methodology for a clock signal with tunable duty cycle. A description of clock tree is also given in this section. Finally, Section \ref{sec:6} concludes the entire paper.

\section{Preliminaries}\label{sec:2}
Before proceeding to substantive parts of this paper, some preliminaries are given in this section for a better explanation of our contributions.

\subsection{Rationale of Pure CRN Design}\label{sec0:0}
Essentially, a mathematical foundation guarantees the safety of our pure CRN design and an RGB(Y) mechanism ensures the oscillation of our clock signal. That is, in 2011, Soloveichik and Winfree proposed in \cite{soloveichik2010dna} that DNA as a universal substrate for chemical kinetics, mathematically confirms that nearly all arbitrary chemical reaction networks (CRNs) could be ``translated'' into DNA strand displacement reactions \cite{zhang2010dynamic,zhang2011dynamic,dnadis,cardelli2013two}, which lays a solid foundation for physical implementation of our proposal. Hence, we are safe in purely designing arbitrary CRNs of target functionality. Moreover, based on the RGB(Y) oscillator, which is first proposed by Jiang Hua in \cite{jiang2010synthesis,jiang2011synchronous}, clock signals of nearly arbitrary duty cycle could be successfully obtained in a general way via our previous work in \cite{ge2015formal}, with an extended method for $1/N$ and novel solutions for both $1/2$ and $N/M$.

\begin{figure}[htbp]
\centering
\includegraphics[width=6.5cm]{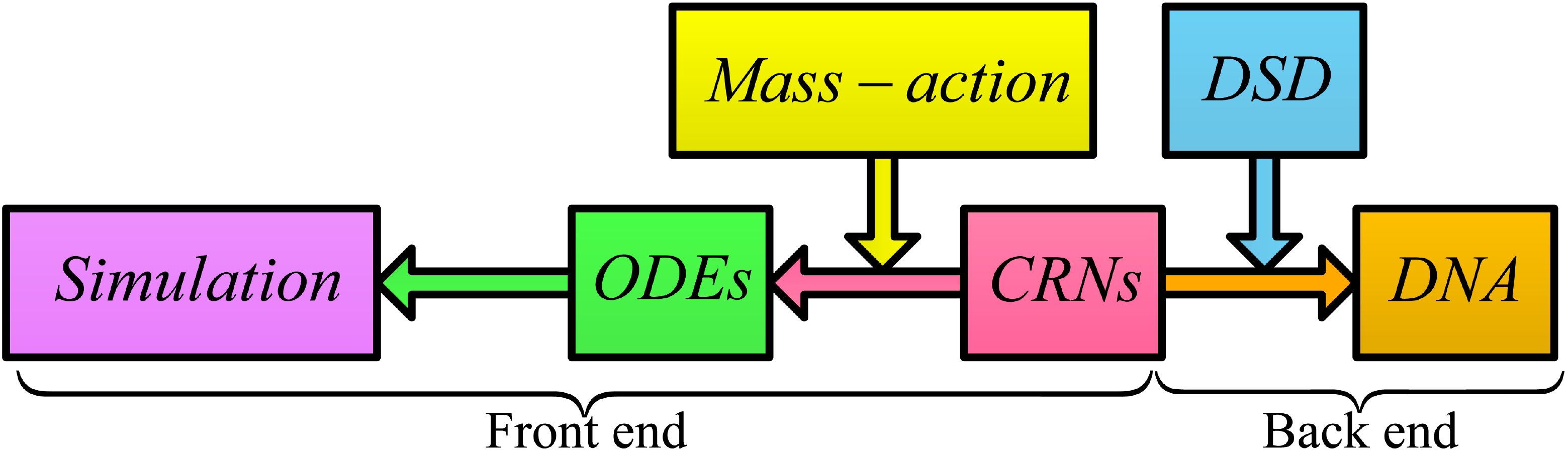}
\caption{A design flow from CRNs to DNA, \textcolor{black}{cited from \cite{llg}.}}\label{fig0:1}
\end{figure}

To better understand what we do, a design flow is illustrated in Fig. \ref{fig0:1}. Feasibilities of the CRNs are guaranteed by numerical results based on the corresponding ODEs, which are obtained via mass-actions \cite{horn1972general,mass}. With the aid of tools such as \textit{Visual DSD} \cite{phillips2009programming} and \textit{NUPACK} \cite{NUPACK}, one could finally translate the designed CRNs into DNA strand displacement reactions with desired clock signal. We view the design of CRNs as ``front end'', whereas view the DNA implementations as ``back end''.

\subsection{Simulation Model}\label{2a}
With the help of ODEs, a simulation model, various dynamic behaviors of certain CRNs could be properly emulated. This ODE-based approach is convenient and can perfectly imitate the time-varying evolution of chemical kinetics. The details of extracting ODEs from a CRN are well-described in \cite{cardelli2008processes}. Key factors are reactants, rate constants, and products.

It is worth noting that although \cite{soloveichik2010dna} requires each chemical reaction has no more than two reactants, in the rest part of this paper, this constraint is ignored for design convenience. Because we can always decompose reactions with more than two reactants into cascaded bimolecular or unimolecular ones. For example, use ``\(A + B \leftrightarrow C\)'' and ``\(C + D \to E + F\)'' to realize ``\(A + B + D \to E + F\)''. Also, there are some CRN-to-DNA translation schemes like \cite{cardelli2013two} that can directly handle higher order reactions. Moreover, as pointed out in \cite{soloveichik2010dna}, the scaled system of lower rate constants and concentrations maintains the same, albeit scaled, behavior. Thus, both the simulated concentration and time are unitless. What's more, for an experimental setup, these systems would be scaled in an appropriate way. For more details, please refer to \cite{soloveichik2010dna,kharam2011binary}.

\begin{figure*}[htbp]
\centering
\subfigure[\textit{Circle map} for duty cycle $1/3$.]{
\label{fig1:a}
\includegraphics[width=4cm]{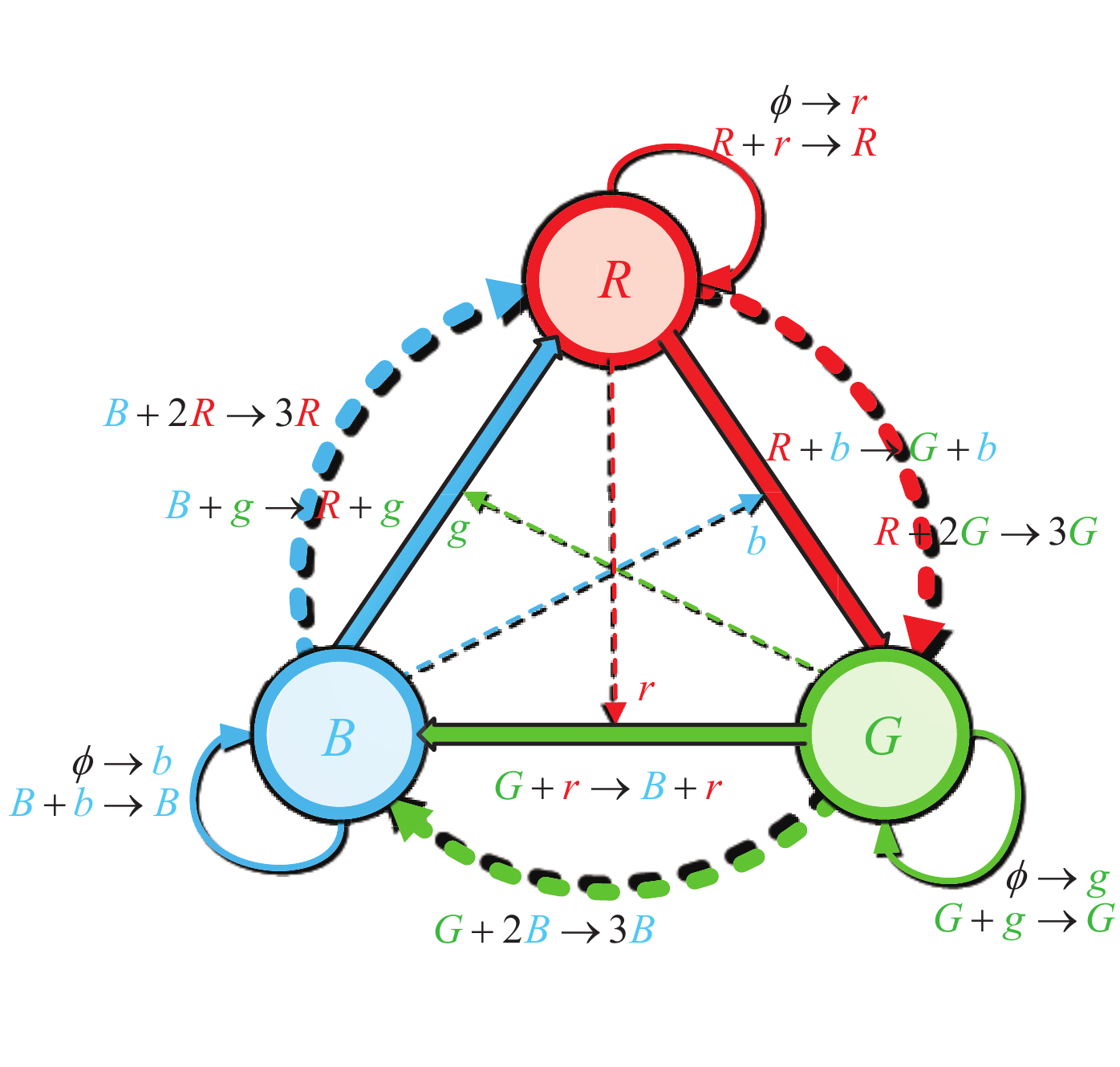}}
\subfigure[\textit{Circle map} for duty cycle $1/2$.]{
\label{fig1:b}
\includegraphics[width=5cm]{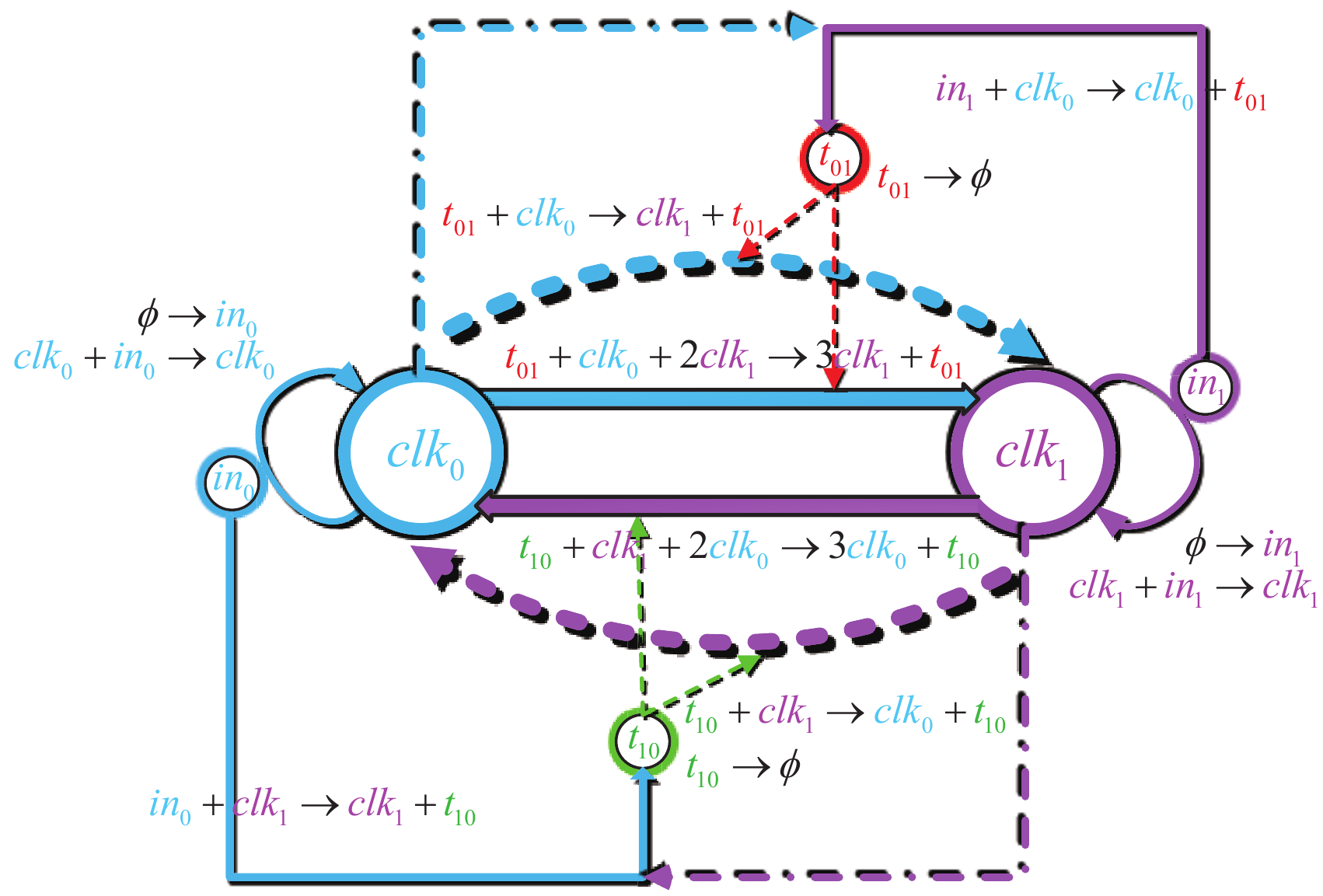}}
\subfigure[Simulation for duty cycle $1/3$.]{
\label{fig1:c}
\includegraphics[width=4cm]{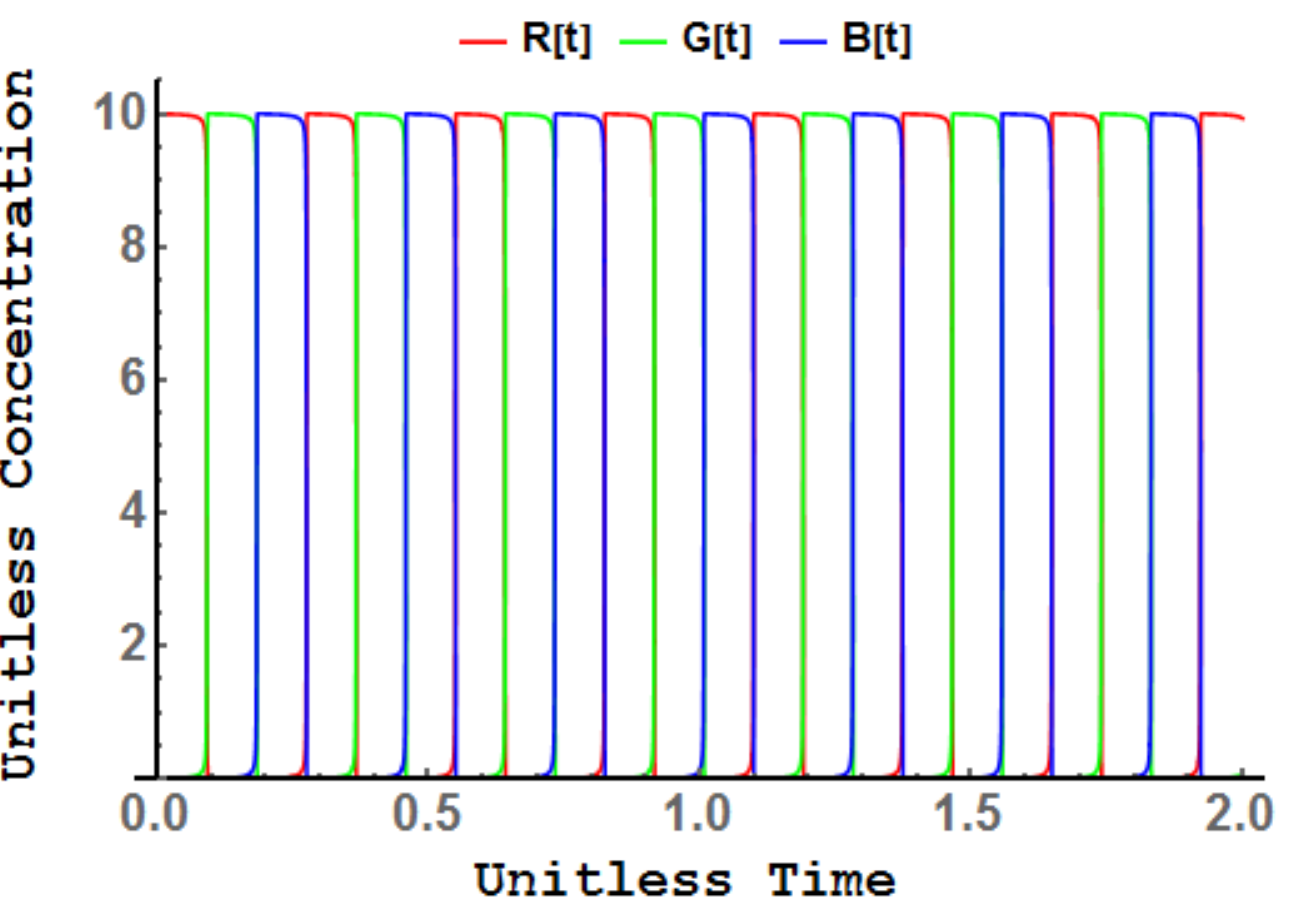}}
\subfigure[Simulation for duty cycle $1/2$.]{
\label{fig1:d}
\includegraphics[width=4cm]{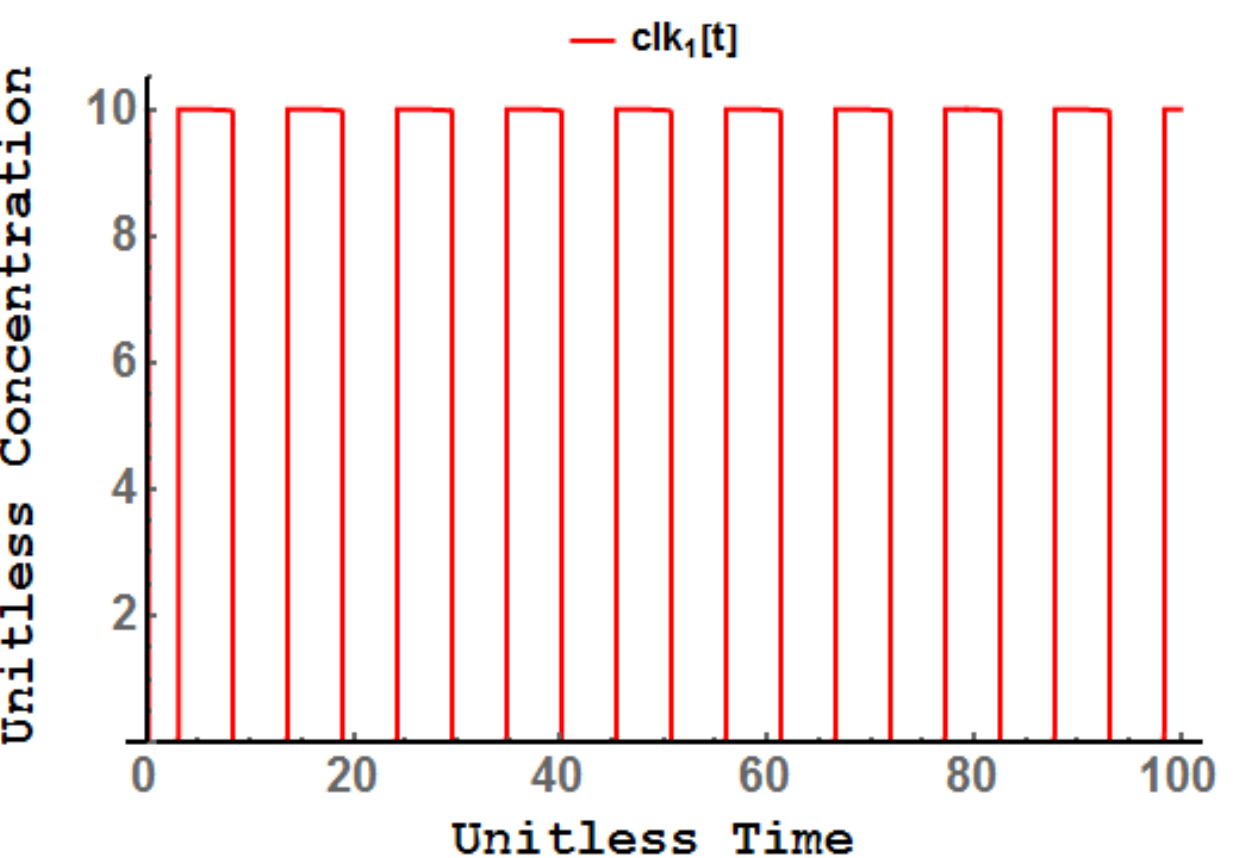}}\\
\subfigure[\textit{Circle map} for duty cycle $3/8$.]{
\label{fig1:e}
\includegraphics[width=8cm]{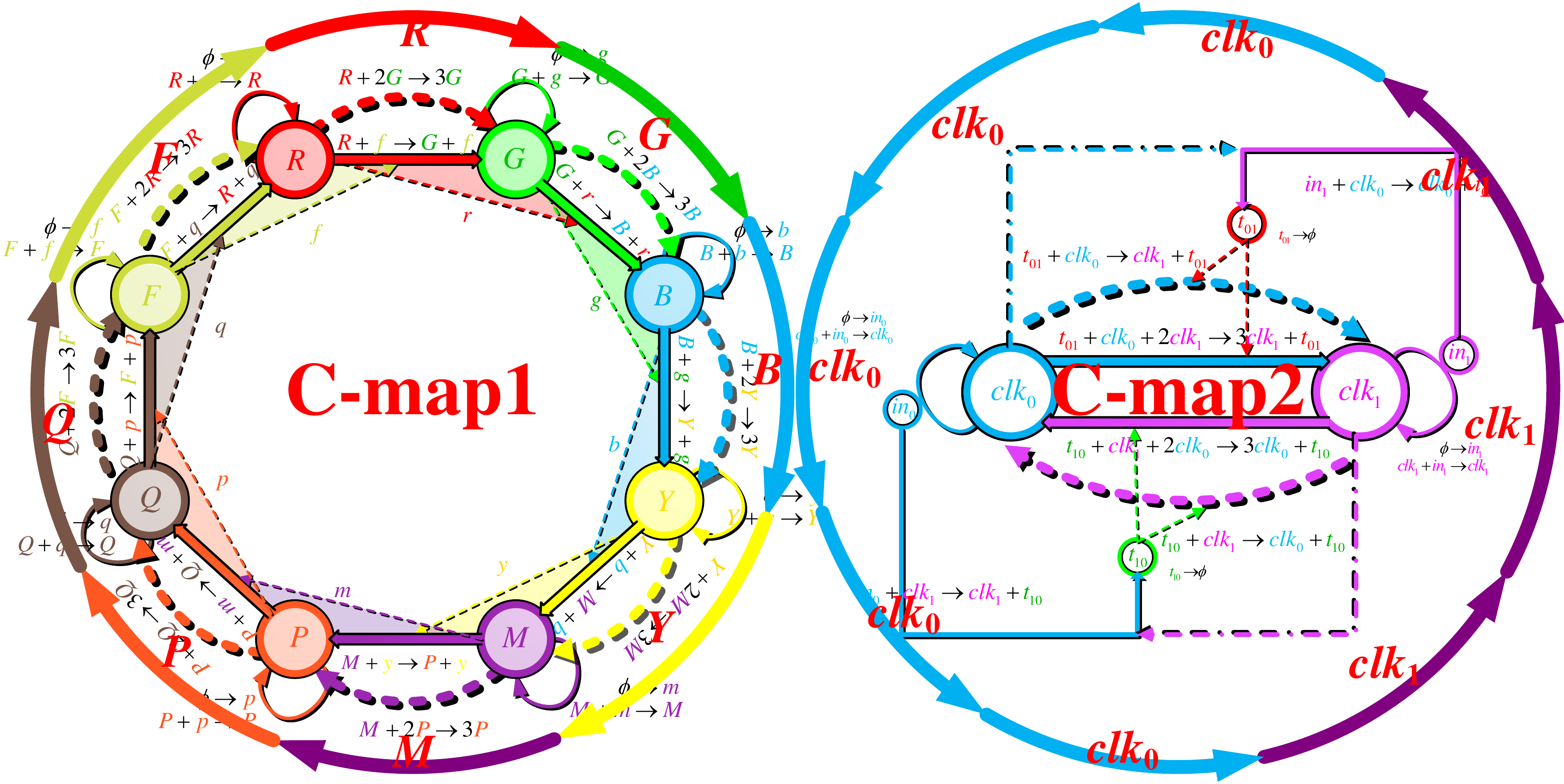}}
\subfigure[Simulation for duty cycle $3/8$.]{
\label{fig1:f}
\includegraphics[width=6.5cm]{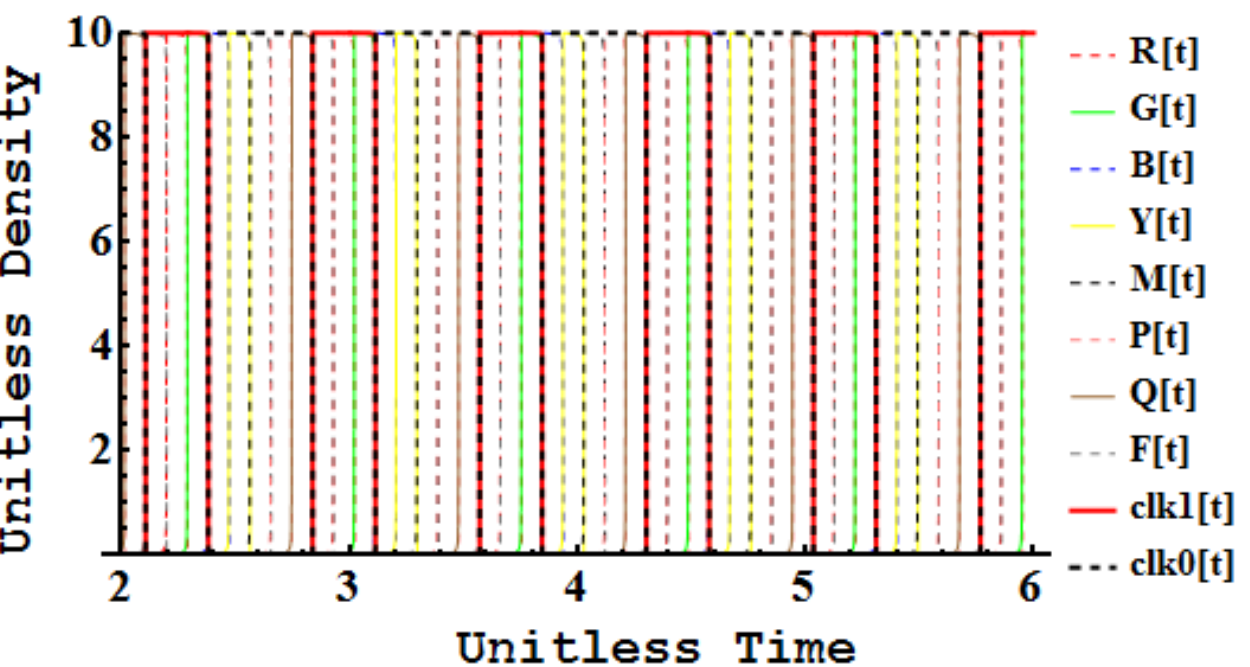}}
\caption{Circle maps and simulation results for three-type implementation methods. \textbf{\textit{(A)}} (a) shows the C-map for $1/3$ duty cycle and can be extended into $1/N$ duty cycle with $N$ nodes. While (b) is the standard C-map for $1/2$ duty cycle with $12$ reactions. The corresponding simulation results are shown in (c) and (d). \textbf{\textit{(B)}} (e) shows the combination of (a) and (b) for $3/8$ duty cycle. Simulation result is illustrated in (f) with the red curve of $clk_{1}$ for $3/8$. With chemical reactions a little fine-tuned, (e) could also realize $1/8$, $2/8$, $4/8$, $5/8$, $6/8$, $7/8$, as well as the extension of $M/N$.}\label{fig1}
\end{figure*}

\subsection{A Quick Review of Our Previous Work}\label{2b}
A \textit{(circle-map)}-aided (C-map) method was proposed in our previous work \cite{ge2015formal}, briefly illustrated in Algorithm \ref{arg1}. We use an $N$-phase oscillator to realize a $1/N$ ($N > 2$) clock signal. This totally requires $4N$ chemical reactions. However, a $1/2$ clock signal is a different case. Its implementation method differs from that of $1/N$ ($N > 2$) and $12$ chemical reactions are in demand. While for an arbitrary $M/N$ clock signal, we need to combine the aforementioned two methods. This $M/N$ clock signal calls for in total $(12+4N)$ chemical reactions with a little changes in $4$ reactions.

\begin{algorithm}
\caption{Implementation of $1/N$ ($N\!\!>\!\!2$), $1/2$, and $M/N$.}\label{arg1}
\begin{algorithmic}[1]
\REQUIRE A system of oscillators and a circle-map (C-map).
\IF {Synthesize $1/N$ ($N > 2$) duty cycle clock signal}
\STATE An $N$-phase oscillator is in demand.
\STATE Due to the given C-map, especially for its arrows, map all the four-type reactions: ``absence'', ``phase signal'', ``threshold'' and ``main power'' reactions.
\STATE \textcolor{black}{Extending} the \textbf{RGB} oscillation mechanism, this implementation calls for $4N$ chemical reactions.
\ELSIF {Engineer a $1/2$ clock signal with a modified C-map}
\STATE To trigger a transference between two phase signals.
\STATE Molecule \({t_{10}}\) and \({t_{01}}\) are introduced to instruct the transference between \(cl{k_1}\) and \(cl{k_0}\).
\STATE Require $12$ chemical reactions.
\ENDIF
\IF{Construct an $M/N$ clock signal}
\STATE Combine both $1/N$ ($N\!\!>\!\!2$) and $1/2$ clock signal implementation. Only $4$ reactions need a little fine-tuning.
\STATE Use different two phase signals of an $N$-phase oscillator to control the transference of $1/2$ clock signal.
\STATE Call for ($12+4N$) reactions.
\ENDIF
\end{algorithmic}
\end{algorithm}

\subsubsection{\textbf{Chemical Reactions for RGB Oscillation}} In our previous proposal, totally four-type reactions are required for RGB oscillation mechanism, named after their functionality: ``absence indicators'', ``phase signals'', ``threshold reactions'' and ``main power reactions''. Concrete chemical reactions are given in Table \ref{t1} in Appendix \ref{sec:app}. Note that an absence indicator is logically complementary to a phase signal, which means in a system, if there exists an absence indicator, then there is no its corresponding phase signal and vice versa.

\subsubsection{\textbf{Chemical Reactions for 1/2 Duty Cycle}} Employing the aforementioned four-type reactions, the implementation of $1/2$ duty cycle calls for $12$ reactions. Introducing two instructors, $t_{10}$ and $t_{01}$, the oscillation mechanism hence could be successfully established between $clk_{0}$ and $clk_{1}$. Concrete chemical reactions are given in Table \ref{t2} in Appendix \ref{sec:app}.

All the C-maps of those three-type implementation methods are shown in Fig. \ref{fig1}, as well as the corresponding simulation results. For more details, please refer to \cite{ge2015formal}.

\section{Clock Design Inspirations from Gears}\label{sec:3}
\textcolor{black}{In this section, after announcing our goal is to design a tunable clock signal, several bijective concepts between gear systems and clock signals are provided. Through analyzing the operating mechanism of gears, several useful instructions are accumulated.}

\subsection{A Tunable Clock Signal}
 To put it simple, our goal is to design a tunable clock signal in a CRN level. In synthetic biology, a good tunable clock signal means parameters, including frequency, period and duty cycle as well as amplitude, could be freely tweaked for various special uses. Since different amplitudes could be easily realized by employing distinctive values of molecule concentrations, our study does not cover more about this simple issue. This paper mainly focuses on the duty cycle implementation. For frequency alteration implementation, please refer to Part II.

\begin{Rem}\label{rem}
It is interesting to note that amplitude nearly takes no place for clock signals in traditional electronics, people only focus on its comparing results with the reference level, hence this kind of comparison will produce a high or low voltage. However, concentrations actually matter a lot in synthetic biology. And in our proposal, the amplitude owns a close relationship with concentration. Therefore, we care about the amplitude of a clock signal.
\end{Rem}

\subsection{\textcolor{black}{Analogy between Gear Systems and Clock Trees}}\label{subsec:3a}
Interestingly, C-maps are found to look like a gear. Different phase nodes look like gear teeth. What's more, the process of a clock signal finishes a time period, just like a gear rotates a cycle at a uniform speed. Thus, an analogy between C-maps, or rather CRNs, and a gear system is direct.

\subsubsection{\textbf{Main Parameters}}
For a single gear, parameters in Table \ref{table00} should be paid attention to. Among all the parameters, the top three are related to the size of a gear. While the bottom four are in relation to its rotation. Note that modem $m$ must be the same for all gears, otherwise they would not mesh.
\begin{table}[ht]
\centering
\caption{Some main parameters}
\begin{tabular}{c||c}
\Xhline{1.0pt}
\textbf{Parameter} &  \textbf{Meaning}\\
\hline
\rowcolor{mygray}
$t$ & Number of teeth on the gear\\
$D$ & Pitch circle diameter  \\
\rowcolor{mygray}
$m$ & modem = $D/t$ \\
$\omega$ & Angular velocity\\
\rowcolor{mygray}
$\upsilon$  & Linear velocity on the circle\\
$Rev$  & Rotational speed (unit: $rev/s$)\\
\rowcolor{mygray}
$o$  & rotation direction\\
\Xhline{1.0pt}
\end{tabular}
\label{table00}
\vspace*{4pt}
\end{table}
\subsubsection{\textbf{Bijective Concepts}} Table \ref{table1} offers a straight bijective concepts between a gear system and a clock tree. \textcolor{black}{We assume the value of torque to drive a gear, is equal to that of transmission energy in a clock tree.} The number of gear teeth has a strong linear relationship with that of phase nodes. So does the relationship between diameter of a gear and length of a clock period, or rather \(D \propto T\). What's more, the angular velocity of a gear remains the same relationship in physics with a clock time period $T$.
\begin{table}[htbp]
\centering
\caption{Mapping from a gear system to a clock tree}
\begin{tabular}{c||c}
\Xhline{1.0pt}
\textbf{Gear system} &  \textbf{Clock tree}\\
\hline
\rowcolor{mygray}
Torque &  Transmission energy \\
Number of gear teeth & Number of phase nodes  \\
\rowcolor{mygray}
Diameter of a gear  & Length of a clock period \\
Angular velocity & \(2\pi\)/clock period\\
\Xhline{1.0pt}
\textbf{Gear} &  \textbf{Clock signal}\\
\hline
\rowcolor{mygray}
Dedendum  & Amplitude\\
Rotation period & Time period \\
\rowcolor{mygray}
Color distribution of the rack & Duty cycle\\
\Xhline{1.0pt}
\end{tabular}
\label{table1}
\vspace*{4pt}
\end{table}

\begin{Rem}\label{rem:2}
As shown in the bottom of Table \ref{table1}, a certain gear could also map with a clock signal in at least three aspects. Since the rotation period of a gear is exactly the time period of a clock signal, the amplitude could be directly reflected from the dedendum of a gear, which means the height of gear teeth is exactly the amplitude of a signal. When it comes to duty cycle of a clock signal, it could be tweaked via \textcolor{black}{different painted schemes} of a certain pinion ``translated'' into the connected rack.
\end{Rem}

\subsubsection{\textbf{Length of a Clock Time Period}} \textcolor{black}{First, we offer a basic idea to define a clock time period of the synthesized chemical clock signal. Both technical measurements and definitions are included. Then, data analysis is conducted to verify the correctness of our clock implementation and definition.}
\paragraph{Basic Idea} Extract data from \textit{Mathematica}, regenerate the curve with \textit{Matlab}, and confirm the specific value of the peak of turning-up and bottom of turning-down. More specifically, use \textit{Case} in \cite{pl} or \textit{Flatten} to extract data in \textit{Mathematica}, then regenerate them in \textit{Matlab}. Finally, utilize the \textit{datatip} in \textit{Matlab} to confirm the wanted value.
\begin{figure}[htbp]
\centering
\includegraphics[width=0.8\linewidth]{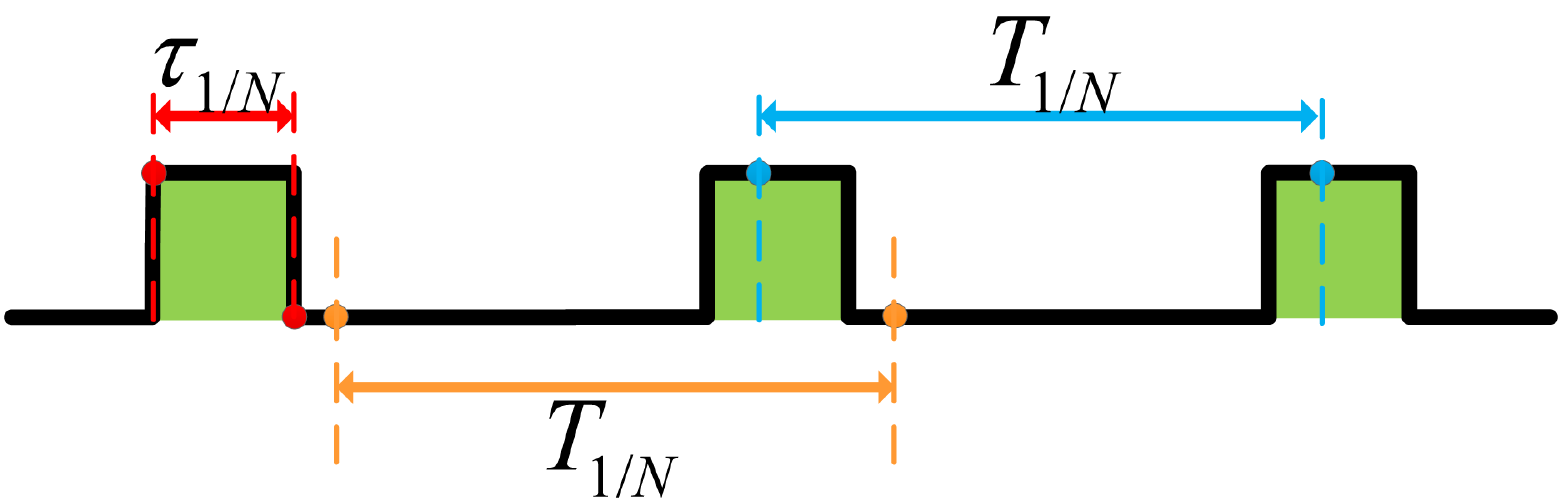}
\caption{Simple model for measuring $T_{1{\rm{/}}N}$ and $\tau_{1{\rm{/}}N}$.}\label{ex}
\end{figure}

Besides, recall that our previous work mainly presented three approaches to construct a clock signal, essentially the approach of $1/2$ differs from that of $1/N$ ($N > 2$), while the clock signal of $M/N$ has the same clock period of $1/N$. Therefore, considering different implementation methods of a clock signal may result in various magnitudes in the length of a clock period, we mainly measure the length of clock time period for $1/2$ and $1/N$.
\begin{figure*}[htbp]
\centering
\subfigure[Regeneration of $1/3$ with \textit{Matlab}.]{
\label{fig2:a}
\includegraphics[width=0.45\linewidth]{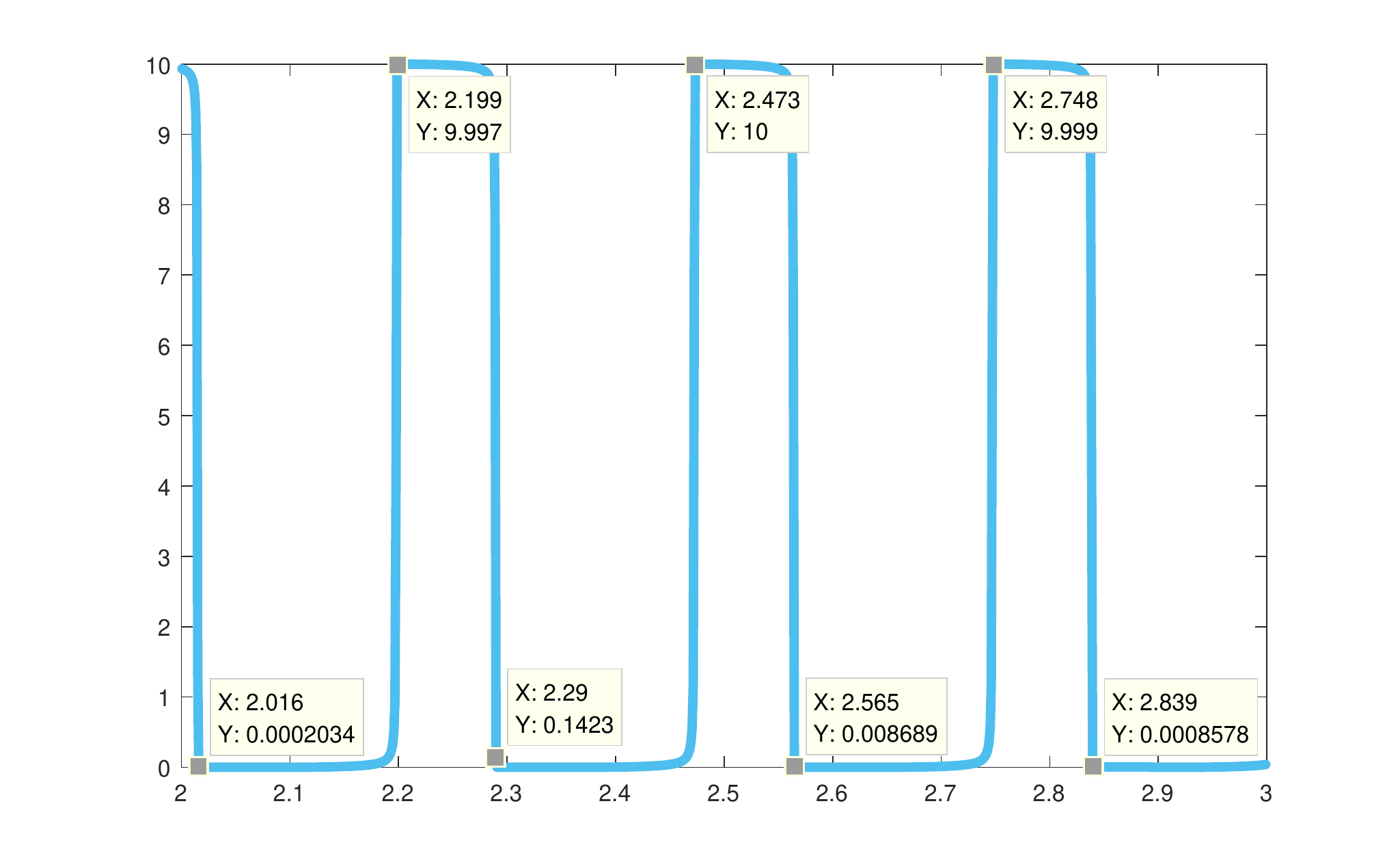}}
\subfigure[Regeneration of $1/2$ with\textit{ Matlab}.]{
\label{fig2:b}
\includegraphics[width=0.45\linewidth]{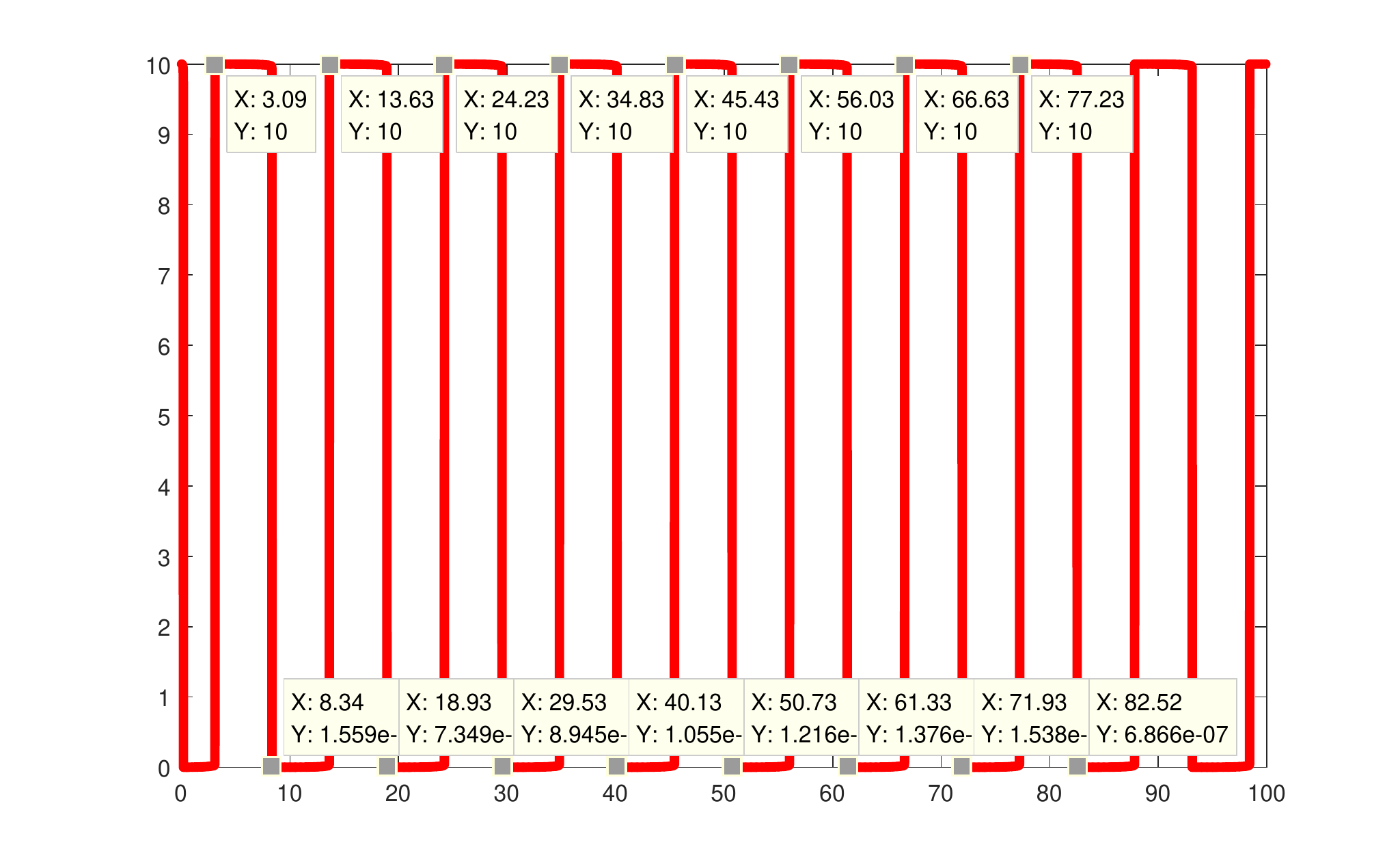}}
\subfigure[Regeneration of $1/5$ with \textit{Matlab}.]{
\label{fig2:c}
\includegraphics[width=0.45\linewidth]{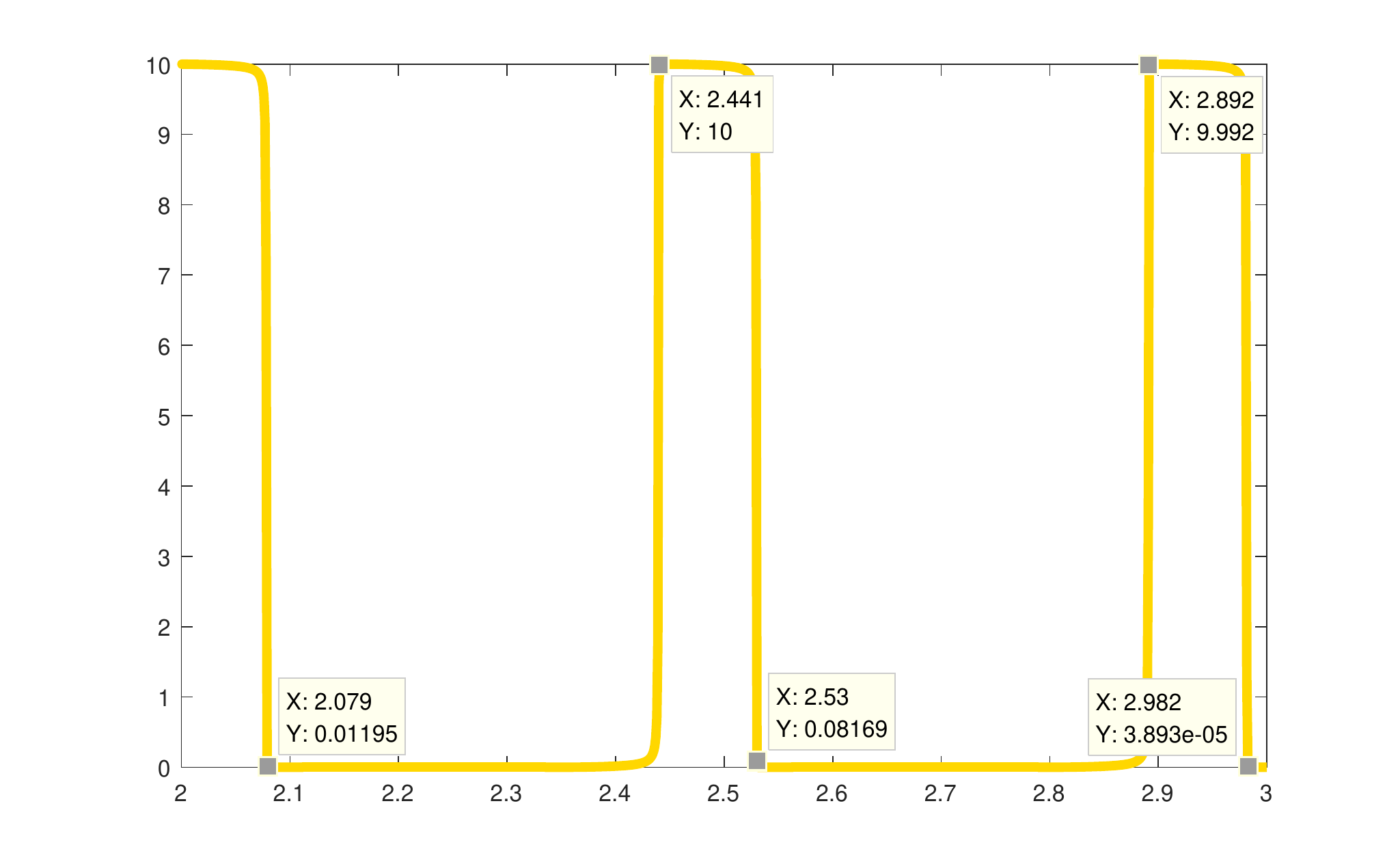}}
\subfigure[Regeneration of $1/15$ with\textit{ Matlab}.]{
\label{fig2:d}
\includegraphics[width=0.45\linewidth]{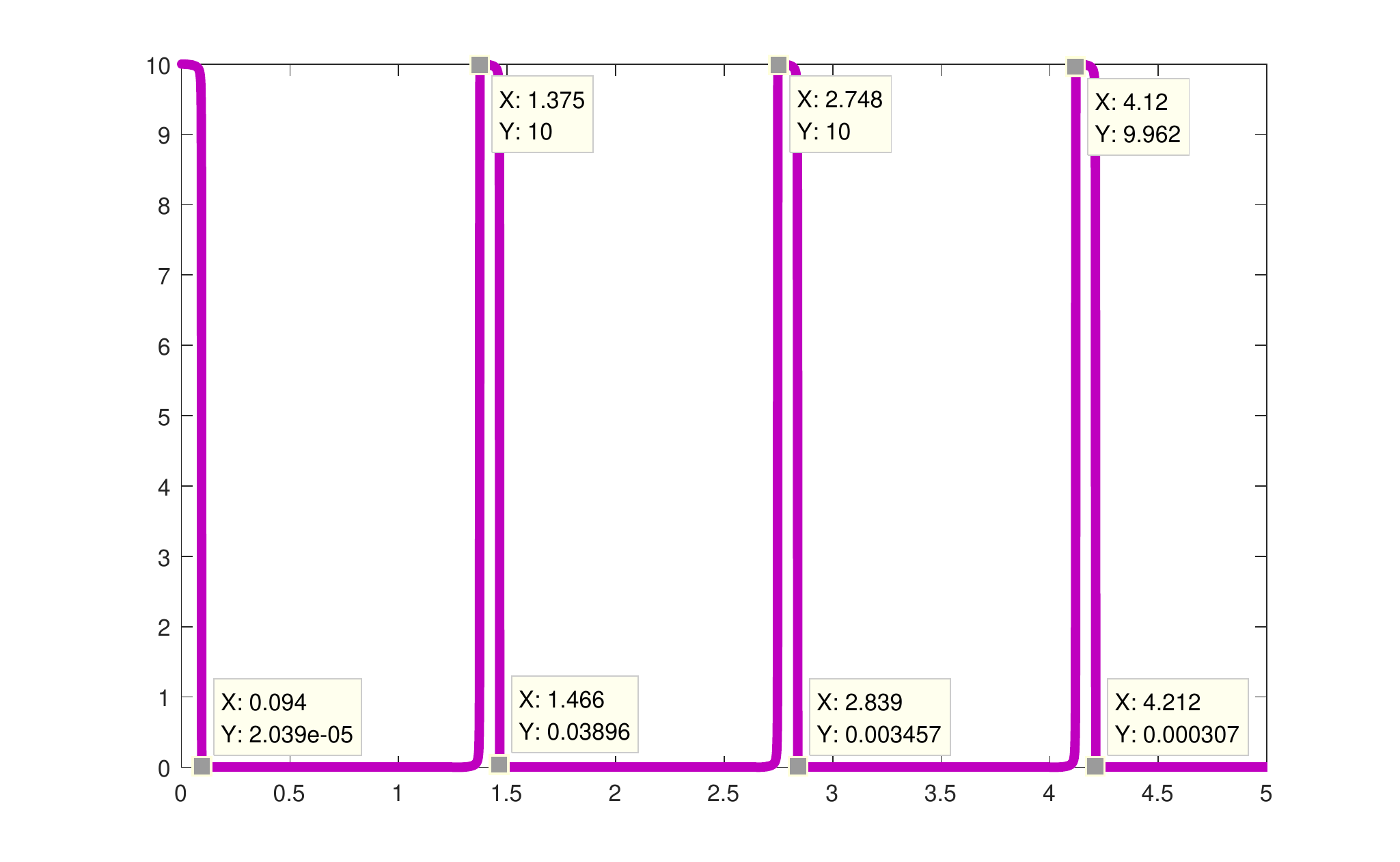}}
\caption{Regenerate data with \textit{datatip} in \textit{Matlab}. The whole period of $1/3$ is $0.275$ unitless time while $1/2$ is $10.6$, $1/5$ is $0.4513$, and $1/15$ is $1.373$.}\label{fig2}
\end{figure*}

\begin{Def}\label{def0:1}
As shown in Fig. \ref{ex}, the length of a clock period is defined as the distance of two neighboring points in the same location of a period signal, top or bottom, denoted $T_{1{\rm{/}}N}$ (colored blue or orange). While the existing time of a phase signal indicates how long a phase signal exists in a cycle of $T_{1{\rm{/}}N}$, or when its concentration is the highest level. Hence $\tau_{1{\rm{/}}N}$ is the distance of two neighboring points that a bottom value minus a previous peak one (colored red). Indeed, \(\hat \tau _{1{\rm{/}}N}\) is the evaluated existing time of a phase signal, with a definition of $T_{1{\rm{/}}N} / N$. All of them contribute to build a better understanding about the gear size in what follows.
\end{Def}

\begin{Rem}\label{rem:3}
Theoretically, different implementation methods lead to different lengths of a clock period, hence we mainly measure the clock period of $1/2$ and $1/N$ ($N > 2$). Due to the implementation method of $1/N$, different phase signals should theoretically have the same existing time, namely $\tau_{1{\rm{/}}N}$. For convenience, clock signals of $ 1/3 $, $ 1/5 $, $ 1/15 $, and $1/2$ are measured.
\end{Rem}

\paragraph{Data Analysis} Processing the data shown in Fig. \ref{fig2}, all the parameters we care most are listed in Table \ref{table1:1}. Carefully observe a single phase existing time of $1/2$ and $1/N$ duty cycle, $\tau_{1{\rm{/}}2}$ is really different from others. While $\tau_{1{\rm{/}}3}$, $\tau_{1{\rm{/}}5}$, and $\tau_{1{\rm{/}}15}$ are nearly the same. This result proves that various implementation methods lead to different $\tau_{1{\rm{/}}N}$. Due to the high value of $T_{1{\rm{/}}2}$ and the rule of \(t \propto D \propto T\), the diameter of a gear representing $1/2$ duty cycle is much bigger than that of $1/3$, $ 1/5 $, and $ 1/15 $ duty cycle.
\begin{table}[htbp]
\centering
\caption{Measurements of $1/2$, $ 1/3 $, $ 1/5 $, and $ 1/15 $}
\begin{tabular}{c||c|c|c|c}
\Xhline{1.0pt}
\textbf{Clock signal}&  $T_{1{\rm{/}}N}$ & $\tau_{1{\rm{/}}N}$ &\(\hat \tau _{1{\rm{/}}N}=T_{1{\rm{/}}N}/N\)& \textbf{Ratio}\\
\hline
\rowcolor{mygray}
$ \bm{1/2} $ & $10.6$   & $5.3$     & $5.3$ & $ \bm{37.8182} $\\
$ \bm{1/3} $ & $0.275$  & $0.092$   & $0.09167$  & $\bm{1}$\\
\rowcolor{mygray}
$ \bm{1/5} $ & $0.4513$ & $0.0895 $ & $0.09026$ & $\bm{1.64109}$\\
$ \bm{1/15}$ & $1.373$  & $0.0913 $ & $ 0.09153 $ & $ \bm{4.99273}$\\
\Xhline{1.0pt}
\end{tabular}
\label{table1:1}
\vspace*{4pt}
\end{table}

\begin{Rem}\label{rem:4}
In Table \ref{table1:1}, the data extracted from Fig. \ref{fig2} are coherent to our design goal, which means our previous design method could really realize $1/2$ and $1/N$ duty cycle. Because $\tau_{1{\rm{/}}N}$ times $N$ approximately equals to $T_{1{\rm{/}}N}$ we measure. Take $1/3$ duty cycle as an example, the product of $\tau_{1{\rm{/}}3}$ and $N=3$ equals to $0.27501$, which is approaching to $T_{1{\rm{/}}3}=0.275$. Hence Fig. \ref{fig2:a} is really simulated the clock signal of $1/3$ duty cycle. So are the other clock signals implemented with our methods.
\end{Rem}

\subsubsection{\textbf{Size of a Gear}} Seizing upon the rule of \(t \propto D \propto T\), we denote ``Ratio'' in Table \ref{table1:1} as ``$T_{1{\rm{/}}N}/T_{1{\rm{/}}3}$''. The last column of Table \ref{table1:1} shows the gear size ratio comparing with the $1/3$ clock signal. Therefore, we use the gear of $1/3$ duty cycle as the standard one, with $D=1$ (unitless). Take a gear of $1/5$ as an example, its diameter $D=1.64109$ and so on.

\subsection{Fundamental Paradigms from Gears to Clock Signals}
Totally four types of gears could be utilized to synthesize chemical clock signals.
\subsubsection{\textbf{Simple Gear Train}}

\begin{figure}[htbp]
\centering
\includegraphics[width=0.9\linewidth]{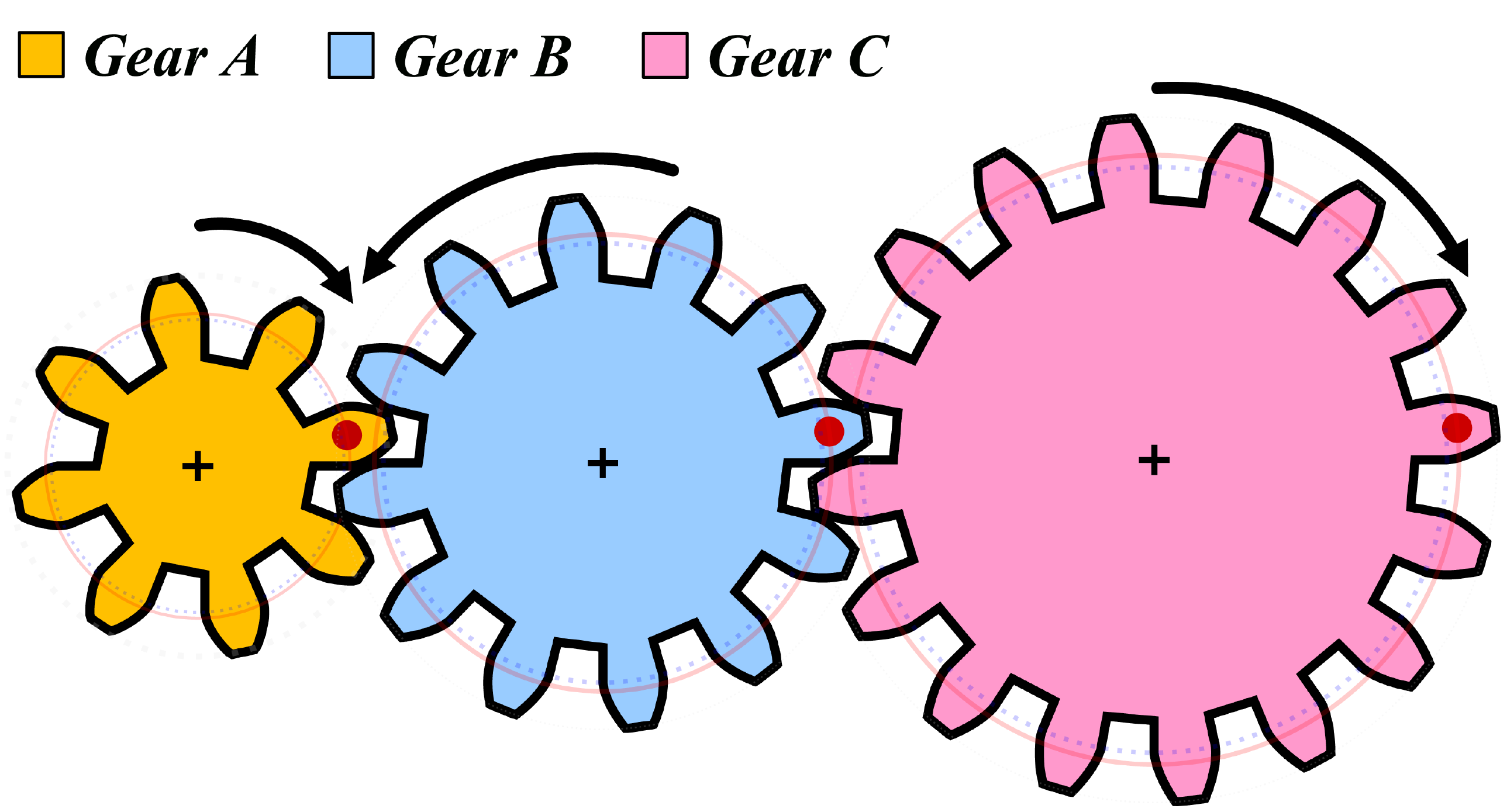}
\caption{A simple gear train.}\label{sim}
\end{figure}

Fig. \ref{sim} typically shows spur gears. Imagine that rotation is from Gear $A$ then $B$ and finally $C$, whereas the direction of this rotation is reversed from one gear to another. Gear $B$, also named an idler gear, has no effect on the gear ratio but merely plays a role in changing the direction of rotation. To mesh with each other, the teeth of all gears must be the same size; hence $m$ for all gears should be identical.

 Velocity is transferred through the implementation shown in Fig. \ref{sim}. All the linear velocity $\upsilon$ must be the same. Gears comply with the rules of Eq. \ref{eq:0}.

\begin{equation}\label{eq:0}
\left\{
 \begin{aligned}
  & {\textstyle{\frac{{\omega _A}{D_A}}{2}}} = {\textstyle{\frac{{\omega _B}{D_B}}{2}}} ={\textstyle{\frac{{\omega _C}{D_C}}{2}}}\\
 \Longrightarrow \; & {\omega _A}{D_A}  = {\omega _B}{D_B} = {\omega _C}{D_C}\\
 \Longrightarrow \; & {\omega _A}m{t_A} = {\omega _B}m{t_B} = {\omega _C}m{t_C}\\
 \Longrightarrow \; & {\omega _A}{t_A}  = {\omega _B}{t_B} = {\omega _C}{t_C}.\\
\end{aligned}
\right.
\end{equation}

For a simple gear train, there exist a lot of physical properties hidden in the following theorems, from which several instructions to model our clock signal of CRNs could be gradually accumulated.

\begin{Thm}\label{thm:0}
A lager-size gear must rotate slower in order to achieve the same $\upsilon$.
\end{Thm}

\begin{IEEEproof}
According to \(\upsilon  = {\textstyle{\frac{{\omega}{D}}{2}}}\) in physics, angular velocity $\omega$ must be lager on condition that the diameter of the gear $D$ is relatively small, hence all the gear share the same linear velocity. Note that rotational speed $Rev$ is physically equal to $\omega$ divided by $2\pi$, namely \(Rev = \omega /2\pi \), therefore $\omega$ is proportional to $Rev$ and smaller gear runs faster.
\end{IEEEproof}

\begin{Cor}\label{cor:0}
If a large gear turns a small gear, the speed increases, and vice versa.
\end{Cor}

\begin{IEEEproof}
Under the circumstances of identical $\upsilon$, Theorem \ref{thm:0} indicates that a smaller gear runs faster than another one between the neighboring two gears. Suppose energy is transferred from a large gear to a smaller one---viewed as an input and an output, respectively---it is obvious that the input gear has a slower $\omega$ while the output faster. Thus, the speed from input to output increases in the aspect of angular velocity.
\end{IEEEproof}

\begin{Thm}\label{thm:1}
For a simple gear train, the relationship among $D$, $t$, and $T$ could be well expressed as \(t \propto D \propto T\).
\end{Thm}

\begin{IEEEproof}
A good mesh calls for the same modem $m$. Because $m=D/t$, therefore $t \propto D$ could be easily obtained. Recall that \(\upsilon  = {\textstyle{\frac{{\omega}{D}}{2}}}\) and \(\omega  = 2\pi /T\), consequently $D \propto T$ when all $\upsilon$ are the same.
\end{IEEEproof}

\begin{figure}
  \centering
  \includegraphics[width=0.8\linewidth]{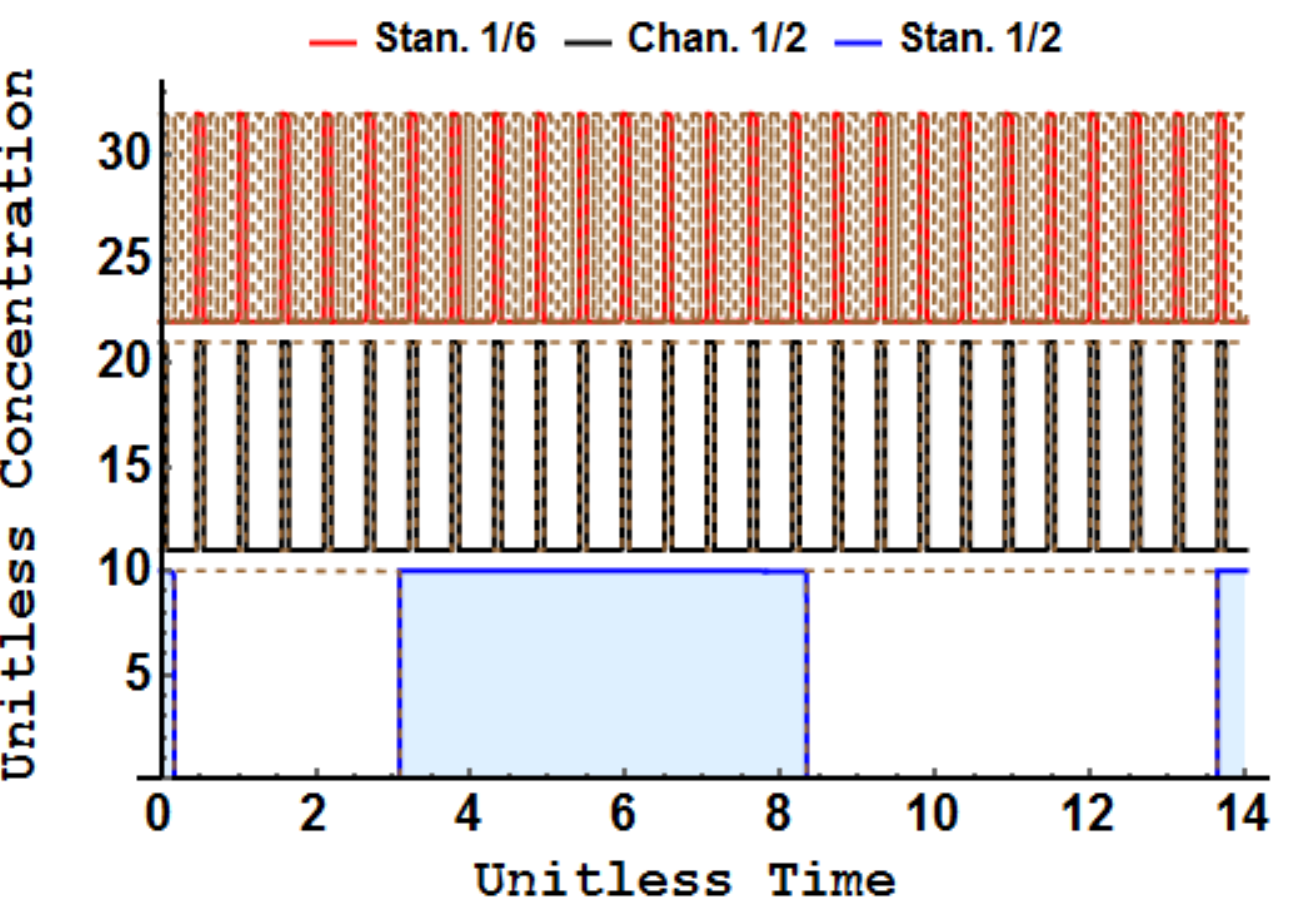}\\
  \caption{CRN simulations for two meshed gears in a simple gear train. The top curve is the standard $1/6$ duty cycle signal, the bottom is the standard $1/2$ duty cycle and the middle is the changed $1/2$ duty cycle signal. }\label{sg1}
\end{figure}

\textbf{\textit{Design Inspirations:}} \textcolor{black}{In practical, we often use a simple gear train to make the time period of a $1/2$ duty cycle clock signal identical with its meshed gear signal. For different gear sizes, although the implementation like Fig. \ref{sim} indicates the same linear velocity but various time periods, we can still change the time period or frequency through an appropriate CRN synthesis. To exemplify this frequency alteration, two meshed gears shown in Fig. \ref{sim} are taken into consideration. We assume that Gear $A$ is a $1/6$ duty cycle clock signal while Gear $B$ represents $1/2$ duty cycle. The simulation of standard $1/6$ and $1/2$ duty cycle are the red and blue curve shown in Fig. \ref{sg1}, and their parameters are in accordance with Table \ref{t2}. We use the first phase of $1/6$ duty cycle clock signal to control the existing time of one phase of $1/2$ clock signal, and the last phase of $1/6$ signal to manipulate the existing time of another phase of $1/2$. The corresponding result is the black curve shown in Fig. \ref{sg1}. This changed $1/2$ duty cycle has the same time period of the standard $1/6$, which differs from its initial time period. One thing should be emphasized is that, the final output is usually meaningless when we use this method to unitize the time period between other kinds of clock signals. Therefore, a simple gear train is often used to unify the time period of a $1/2$ duty cycle clock signal with that of another arbitrary duty cycle. Other kinds of frequency or time period alteration should employ other different gear models.}

\subsubsection{\textbf{Compound Gears}}\label{subcg}
\begin{figure}[htbp]
\centering
\includegraphics[width=0.4\linewidth]{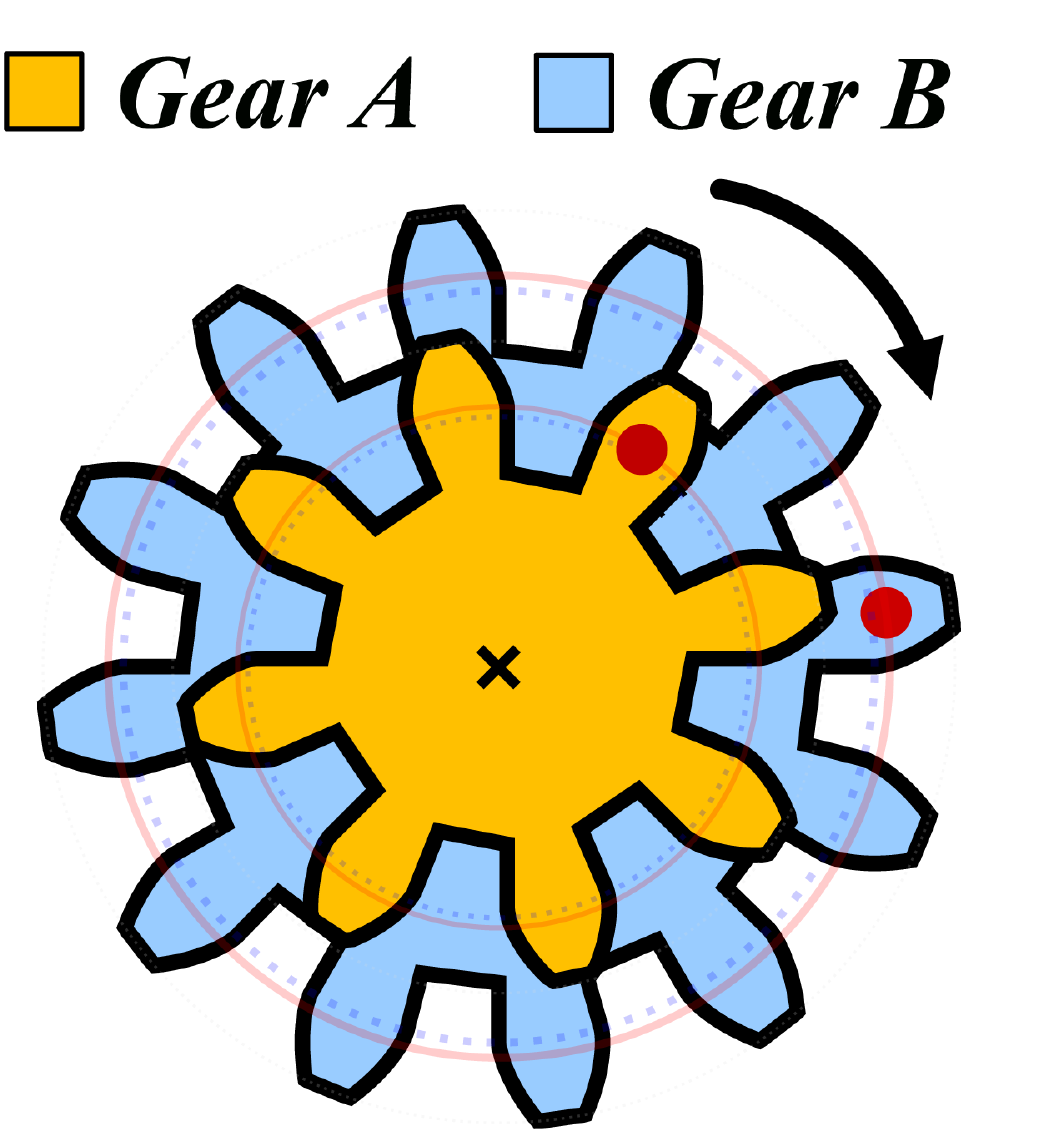}
\caption{Simple model for compound gears.}\label{sim:3}
\end{figure}

Compound gears are widely used in many mechanical devices like engines and workshop machines. As Fig. \ref{sim:3} indicates, compound gears are actually two gears attached to each other and they rotate around the same center. Sometimes compound gears are employed, thus the final gear in a gear train could rotate at a target speed. One of the most important properties for compound gears is illustrated in Theorem \ref{thm:2}.

\begin{Thm}\label{thm:2}
Compound gears always have the same time period $T$.
\end{Thm}

\begin{IEEEproof}
Rotating around the same center, compound gears share the same angular velocity $\omega$. Based on the physical rule of \(\omega  = 2\pi /T\), they also have an identical $T$ but different $\upsilon$ for different $D$.
\end{IEEEproof}

\begin{figure}
  \centering
  \includegraphics[width=0.8\linewidth]{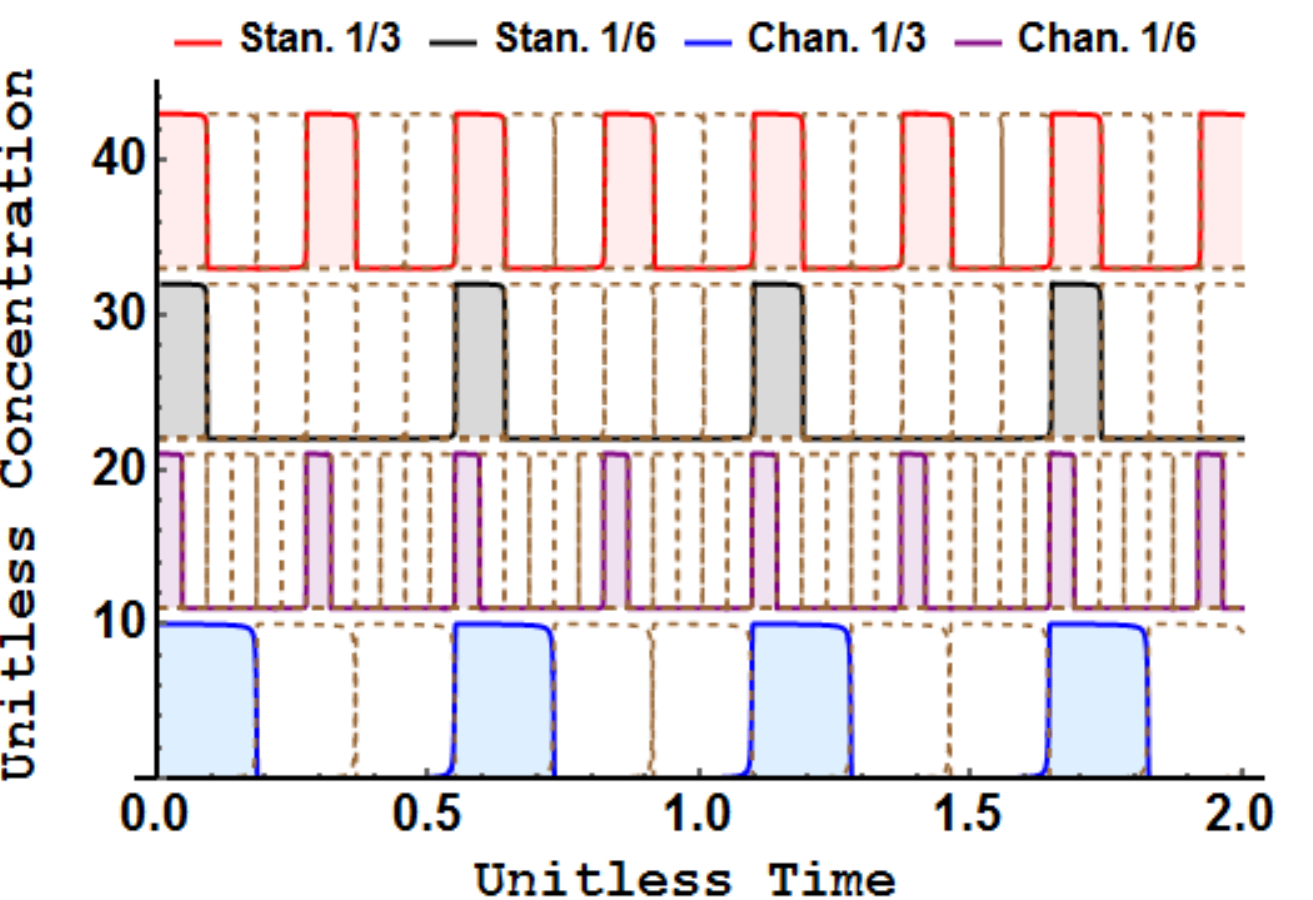}\\
  \caption{CRN simulations for compound gears when purely change rate constants. The top two curves are the standard clock signals of $1/3$ and $1/6$ duty cycle. Others are the changed signals.}\label{cg1}
\end{figure}

\begin{figure}[bpht]
  \centering
  \includegraphics[width=0.8\linewidth]{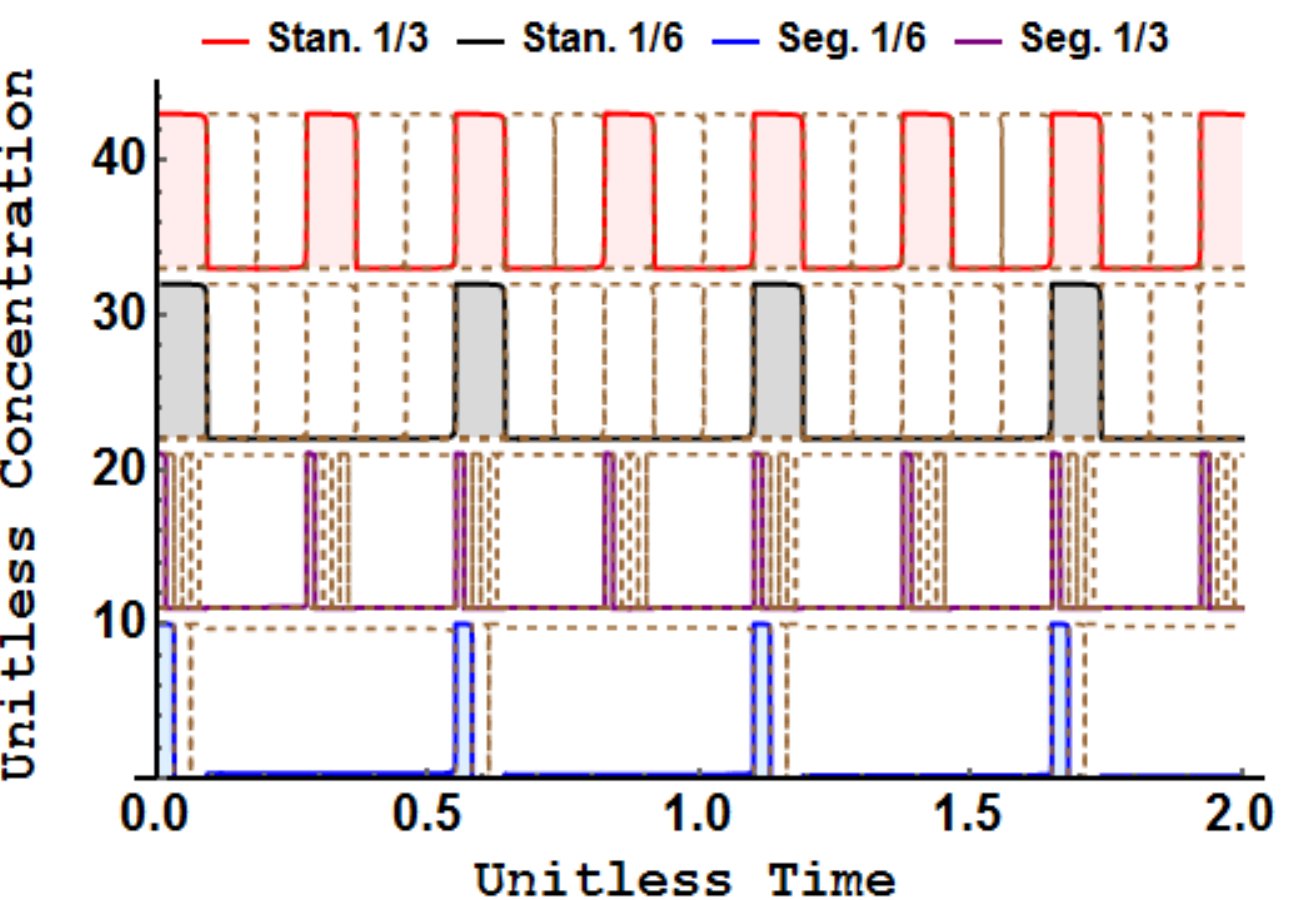}\\
  \caption{CRN simulations for compound gears when segment one phase existing time. The top two curves are the standard clock signals of $1/3$ and $1/6$ duty cycle. Others are the segmented signals.}\label{cg2}
\end{figure}

\textbf{\textit{Design Inspirations:}} \textcolor{black}{This compound gear model is usually used for frequency alteration in terms of CRN synthesis. Two ways are motivated to unitize the time period of two gears through this implementation in Fig. \ref{sim:3}. Assume Gear $A$ is a three-phase signal, while Gear $B$ a six-phase one. \textbf{\textit{(1)}}. Purely change the rate constants of the respective CRNs, especially those constants of threshold and main power reactions. When both Gear $A$ and $B$ take the standard parameters as in Table \ref{t2}, the corresponding results are the top two waves shown in Fig. \ref{cg1}. Note that their time periods are different. For Gear $B$, adopt $1.55$ as the rate constant of threshold reactions and $500$ as that of main power reactions. The changed clock signal colored purple is still $1/6$ duty cycle, but has the same time period as the standard $1/3$ duty cycle of Gear $A$. Similarly, the changed $1/3$ duty cycle clock signal could own the identical time period of the standard $1/6$ one when it adopts $0.5$ as the threshold reactions' rate constant and $26$ as the main power reactions'. The changed two clock signals are shown in the bottom of Fig. \ref{cg1}. \textbf{\textit{(2)}}. Segment a phase signal into pieces and still adopt appropriate rate adjustments. This means we use a phase of an oscillator to control the threshold and main power reactions of another oscillator. The bottom curve of Fig. \ref{cg2} is the simulation when we use one phase of the standard $1/6$ duty cycle clock signal to control the aforementioned two-type reactions of a $1/3$ one. This blue curve is actually the output of the three-phase oscillator, but has the same time period of the standard $1/6$ duty cycle clock signal. Since this CRN synthesis is essentially to segment one phase of $1/6$ into three pieces, the final output is actually a $1/18$ duty cycle. Therefore, this kind of frequency alteration also changes the final duty cycle. A purple curve in Fig. \ref{cg2} is the result of segmenting one phase of $1/3$ duty cycle into six pieces, and it is also a $1/18$ duty cycle. Interestingly, although both the bottom two curves are $1/18$ duty cycle clock signal, their time period are different. Therefore, our further CRN synthesis of frequency alteration in Part II is based on compound gear model. More detailed discussion will also be offered.}


\subsubsection{\textbf{Compound Gear Train}}
\begin{figure}[htbp]
\centering
\includegraphics[width=0.9\linewidth]{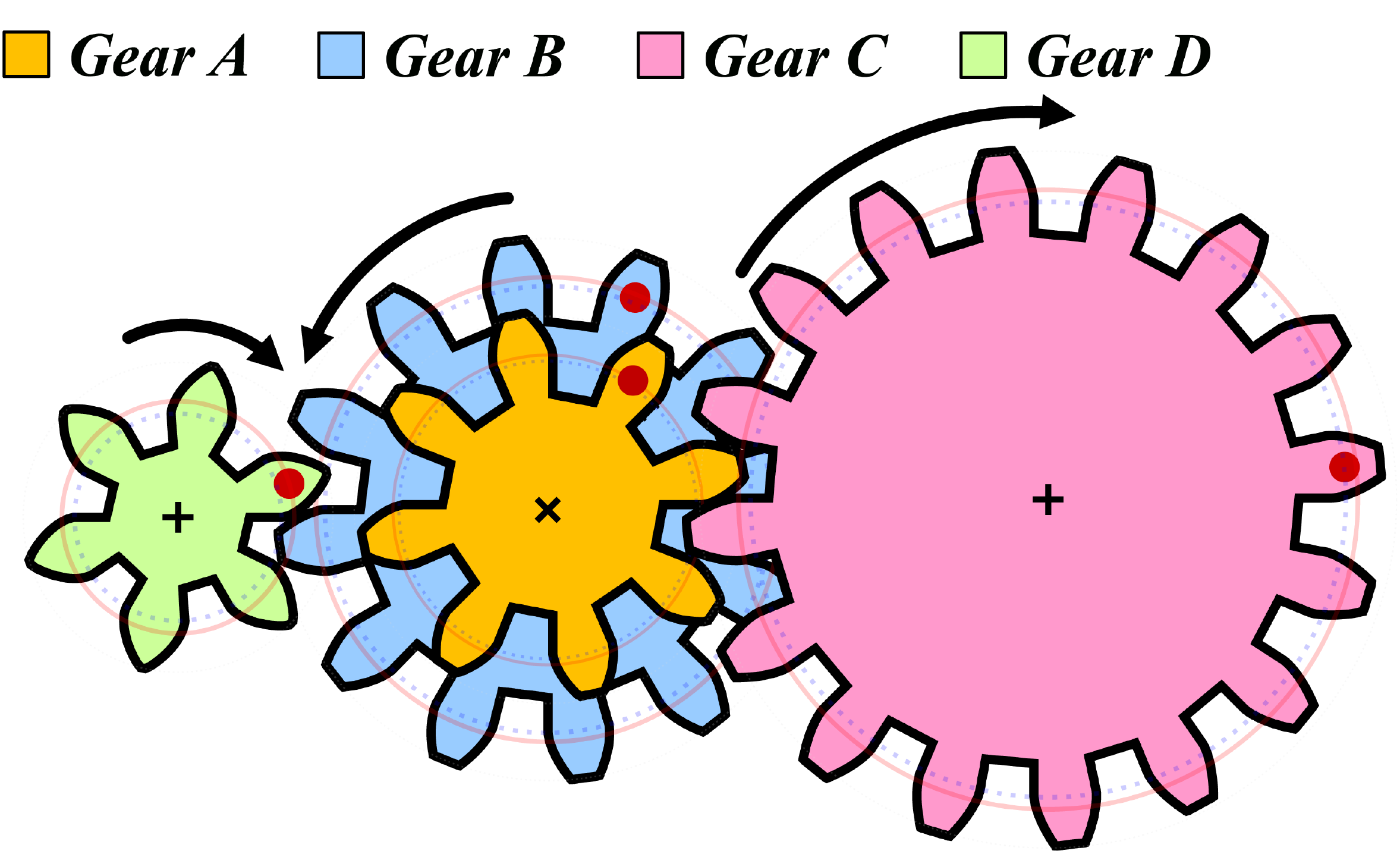}
\caption{Simple model for a compound gear train.}\label{sim:4}
\end{figure}

A compound gear train is an upgraded version of compound gears. As shown in Fig. \ref{sim:4}, suppose that energy is transferred from left to right one by one, both \{Gear $A$, Gear $C$\} and \{Gear $B$, Gear $D$\} compose a pair of simple gear train with the same $\upsilon$, respectively. While \{Gear $A$, Gear $B$\} implements a compound gear with the property of identical $\omega$.

\textbf{\textit{Design Inspirations:}} \textcolor{black}{A compound gear train is usually employed to model a clock tree. Because this implementation is composed of compound gears and a simple gear train, thus the time period of a clock signal could be controlled through the selection of appropriate gear sizes for the meshed gears. Also, various frequency alterations could be realized in this implementation. Due to the page limit, a case study is omitted.}

\subsubsection{\textbf{Rock and Pinion}}
Besides the aforementioned implementations, a gear could also mesh with a linear-toothed part, called a rack, hence produces translation instead of rotation.

\begin{figure}[htbp]
\centering
\includegraphics[width=0.7\linewidth]{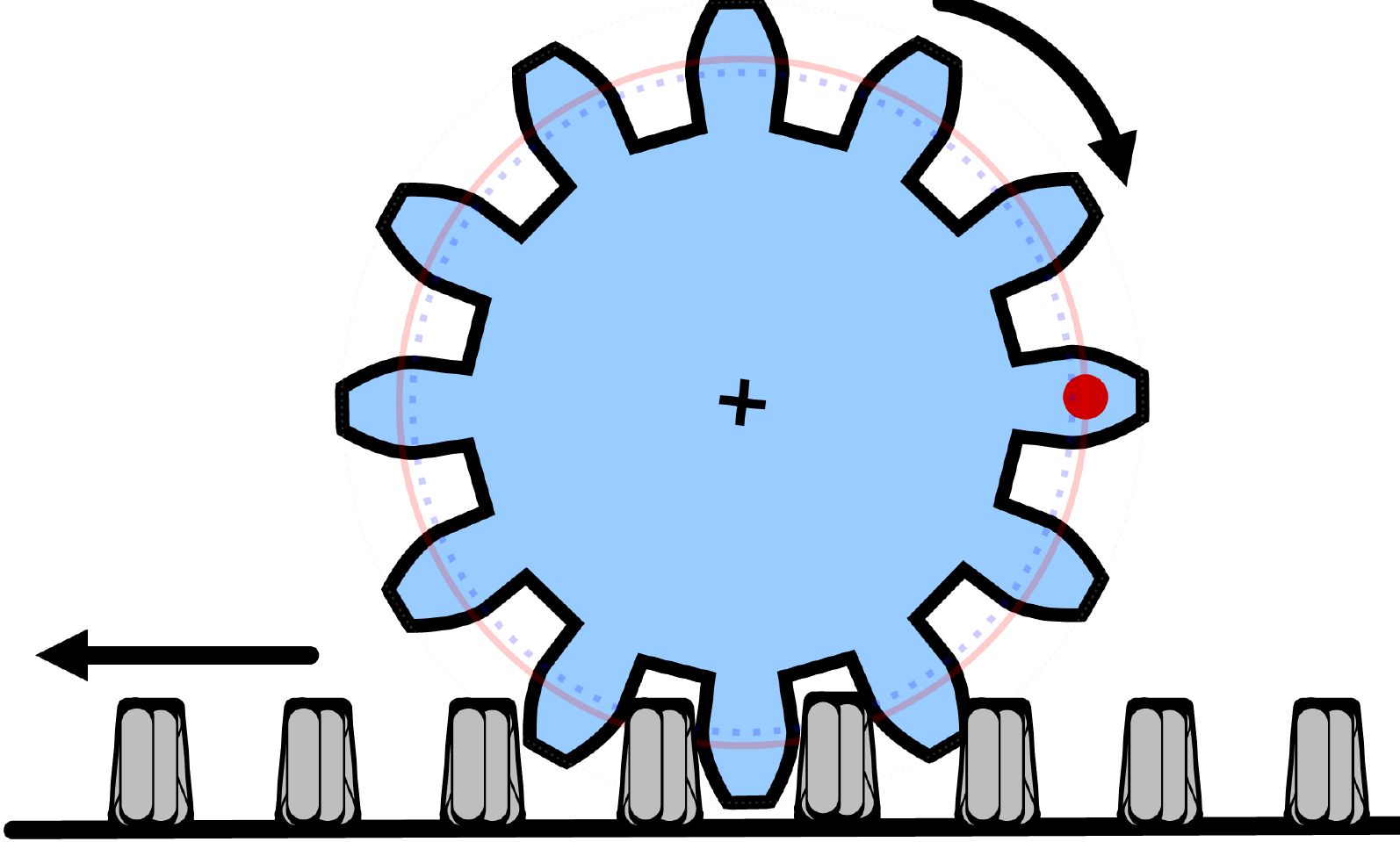}
\caption{Simple model for a rack and pinion gear system.}\label{sim:5}
\end{figure}

According to Fig. \ref{sim:5}, although it is still composed of two gears, this rack and pinion gear system looks quite unusual. Actually, the pinion is a normal round gear while the rack is a straight one with teeth. Therefore, the rack could move the pinion gear only if their respective teeth could mesh well. Moreover, if the pinion rotates in a round, the rack will move in a straight line---another way of describing this phenomenon is to say the ``rotary motion'' is changed to be a ``linear one''.

\textbf{\textit{Design Inspirations:}} Since a ``rotary motion'' could be altered into a linear one, a visual clock signal could be obtained via the implementation of a rack and pinion gear system. Moreover, to reflect the specific duty cycle, the pinion should be colored with two colors. Thus, the rack could well display the duty cycle in a line that we are familiar with.

\section{Gear Models for Clocks with Duty Cycle}\label{sec:4}
\textcolor{black}{This section proposes different operating mechanisms for three-type clock signals, including duty cycles of $1/2$, $1/N$ ($N > 2$), and $M/N$. Moreover, the operation of a clock tree is also elaborated in this section.}

\begin{figure*}[htbp]
\centering
\subfigure[Gear of $1/2$ duty cycle.]{
\label{fig3:a}
\includegraphics[width=4cm]{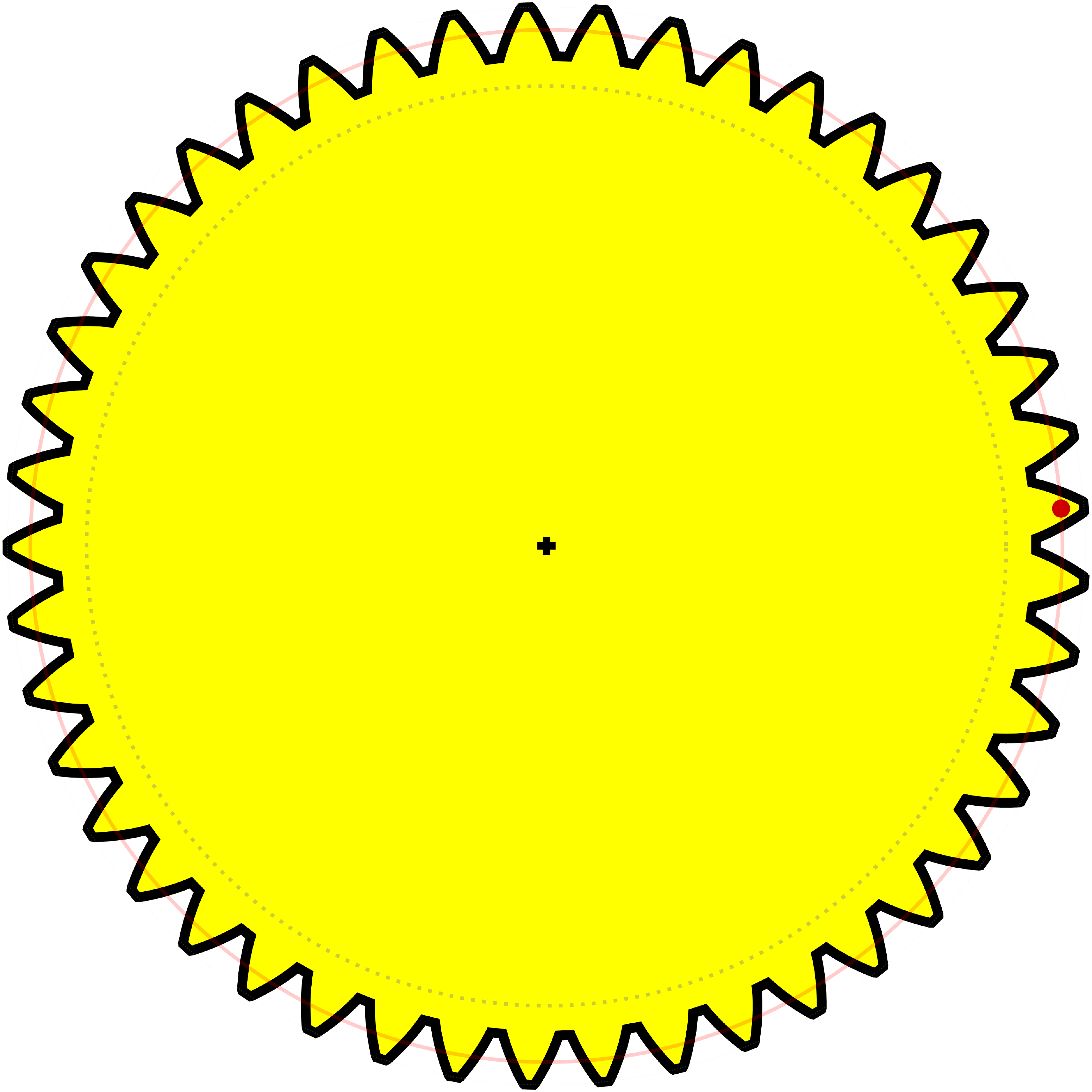}}
\subfigure[Gear of $1/N$ duty cycle.]{
\label{fig3:b}
\includegraphics[width=3.5cm]{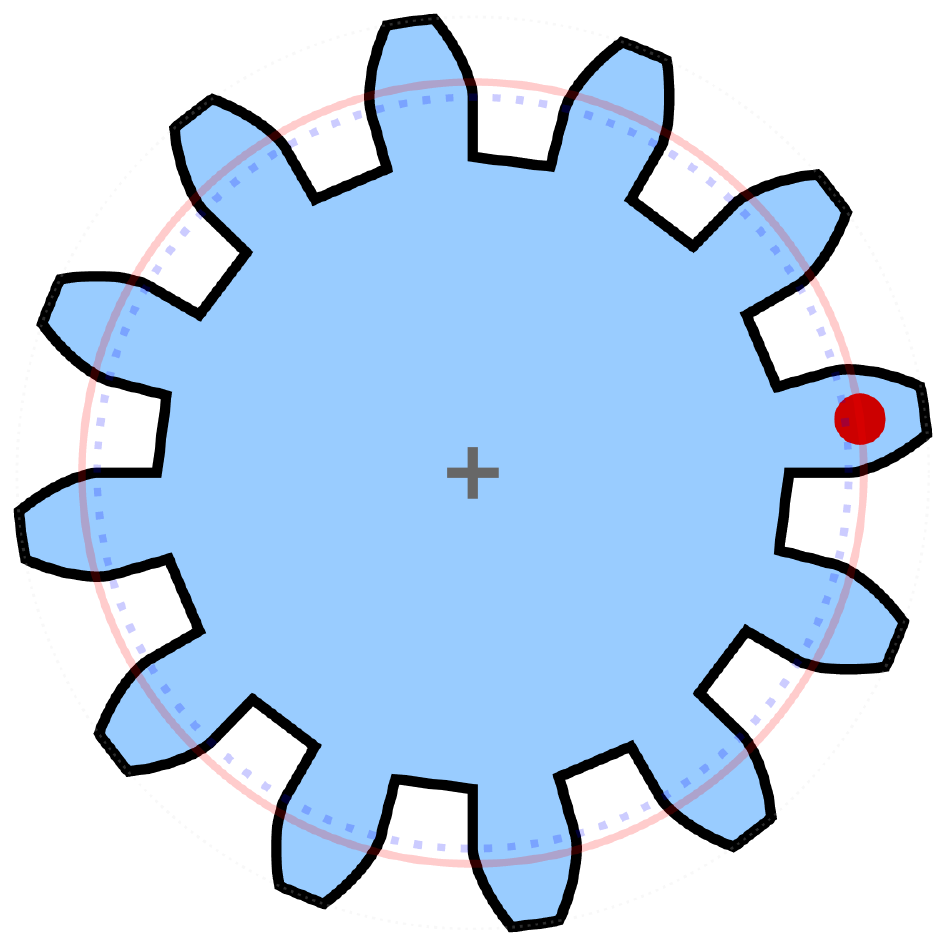}}
\subfigure[Implementation of $M/N$ duty cycle.]{
\label{fig3:c}
\includegraphics[width=0.55\linewidth]{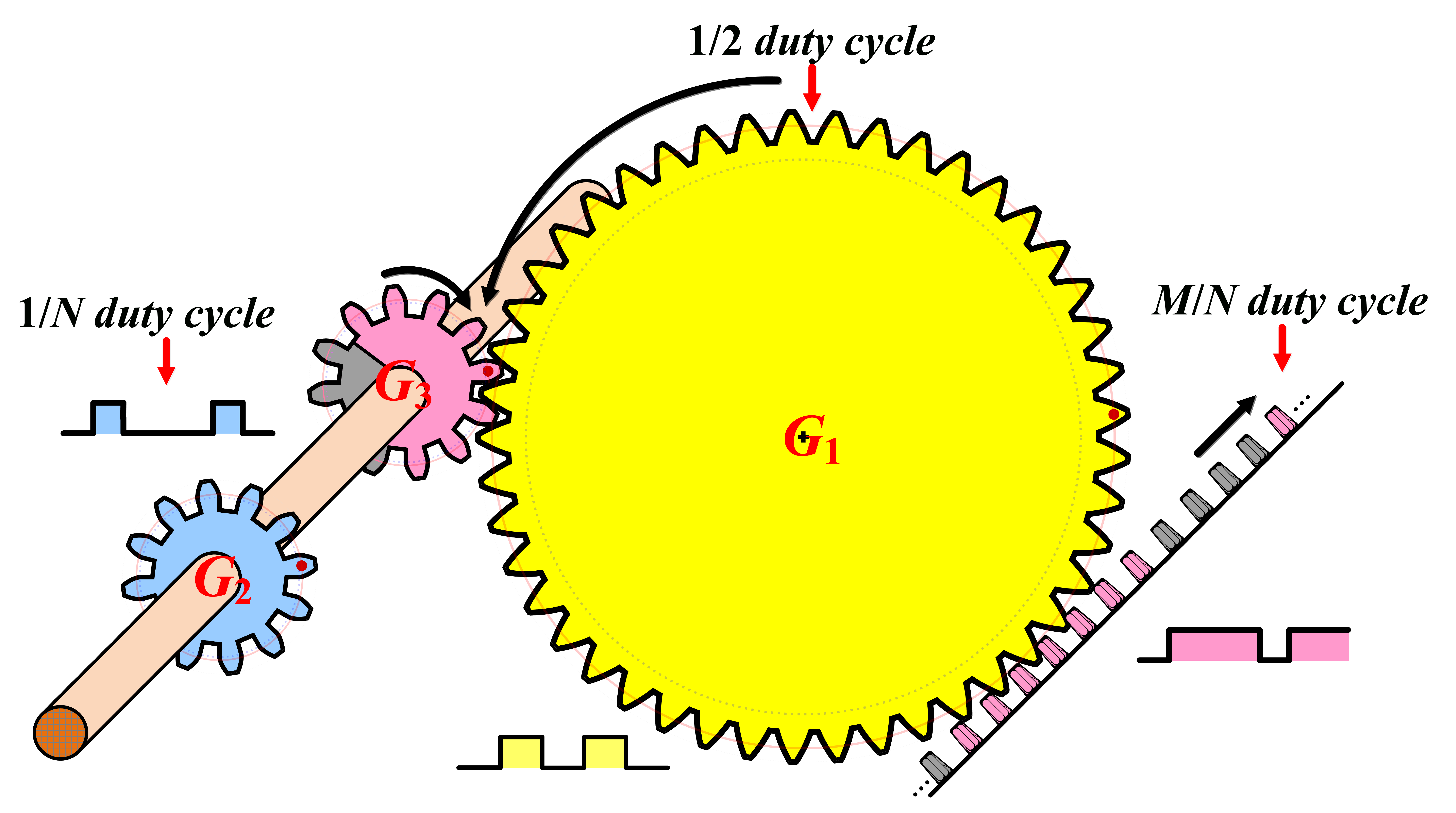}}
\caption{Gear model of $1/2$, $1/N$, and $M/N$ ($N > 2$) duty cycle clock signal. \textbf{\textit{(A)}} (a) shows the gear representing $1/2$ duty cycle. \textbf{\textit{(B)}} (b) shows the gear standing for $1/N$ ($N > 2$) duty cycle. \textbf{\textit{(C)}} (c) displays the gear denoting $M/N$ duty cycle. Where \({G_1}\) represents $1/2$ duty cycle clock signal, \({G_2}\) realizes the functionality of $1/N$ ($N > 2$) clock signal. While \({G_3}\) is an concentric wheel of \({G_2}\), and all its parameters are the same as \({G_2}\). With the combination of \({G_1}\) and a single \({G_2}\), a clock signal of $M/N$ duty cycle could be basically implemented. While \({G_3}\) is an auxiliary to better display the target clock signal.}\label{fig3}
\end{figure*}

\subsection{Operating Mechanism}\label{subsec:3b}
On a chassis of analogy mentioned above, we propose an operating mechanism of our previous clock implementations as follows.

\subsubsection{\textbf{For a Single 1/2 Clock Signal}} In our proposal, a clock signal of $1/2$ duty cycle is denoted by a gear with nearly $227$ ($6 \times 37.8182$) teeth if the standard gear of $1/3$ duty cycle has $6$ teeth. Meshed with other relatively tiny gears of $1/N$ ($N > 2$), the drawn gear of $1/2$ in Fig. \ref{fig3:a} only has $45$ teeth, but actually represents $227$ teeth. It is interesting to note that this yellow gear looks like the C-map shown in Fig. \ref{fig1:b}.

\subsubsection{\textbf{For a Single 1/N Clock Signal}} Each $1/N$ ($N > 2$) clock signal works as a single rotated gear as shown in Fig. \ref{fig3:b}. Although the $1/2$ duty cycle clock signal is also functionally equal to a single rotated gear, its effective diameter is distinctively different because it uses another implementation method. More specifically, the gear size of $1/N$ duty cycle equals to that of $1/2$ only when $N=117$, \textcolor{black}{therefore the gear size of $1/N$ is usually smaller than that of $1/2$ duty cycle if \(N \ll 117\)}. Similarly, this blue gear quite looks like the C-map shown in Fig. \ref{fig1:a}.

\subsubsection{\textbf{For a Clock Signal of M/N Duty Cycle}} In fact, the operating mechanism of $M/N$ duty cycle that we propose works like what Fig. \ref{fig3:c} shows. Both \({G_1}\) and \({G_2}\) are required to implement a clock signal with $M/N$ duty cycle, representing $ 1/2 $ and $ 1/N $ ($N > 2$) duty cycle, respectively. While \({G_3}\), painted with two colors, is used to ``segment'' the denominator $N$.

\begin{Rem}\label{rem:5}
From a standpoint of CRNs, after the implementation of CRNs for both $1/2$ and $1/N$ duty cycle clock signal, an interaction of these two CRNs should be established; thereby two phase signals of $N$ oscillator controlling the transference of $1/2$ clock signal, especially the procedure of its ``threshold'' and ``main power'' reaction.
\end{Rem}

To be specific, based on Remark \ref{rem:5}, the operating mechanism is proposed as follows: \({G_2}\) and \({G_3}\) compose the compound gears as shown in Section \ref{subcg}, sharing the same gear size and velocity. While \({G_1}\) and \({G_3}\) constitute a simple gear chain, transferring the \textcolor{black}{painting} for $M/N$. After that, the final $M/N$ duty cycle clock signal is obtained via the implementation of a rock and pinion, translating the ``rotary motion'' into a linear one. Note that \({G_3}\) is colored by two different colors, one is pink for $M$ sections, and another is grey for $N-M$ sections. The final time period is the same as \({G_2}\), but differs from that of the initial \({G_1}\).

\begin{Rem}\label{rem:6}
When it comes to $M/N$ duty cycle clock signal, the implementation of \({G_1}\) and \({G_3}\) looks like the corresponding C-map in Fig. \ref{fig1:e}. Note that there is no essential distinction between \({G_2}\) and \({G_3}\), because \({G_3}\) is introduced for the convenience of painting.
\end{Rem}

These $3$-type gear models are summarized in Algorithm \ref{arg3}.

\begin{algorithm}
\caption{Gear models of $1/2$, $1/N$ ($N > 2$), and $M/N$.}\label{arg3}
\begin{algorithmic}[1]
\REQUIRE Three-type gears.
\IF {Implement a clock signal of $1/2$ duty cycle.}
\STATE This signal is modeled as a single gear with $227$ teeth.
\ELSIF {Implement a clock signal of $1/N$ duty cycle.}
\STATE Take the gear of $1/3$ with $6$ teeth as a standard. Its diameter is $1$ unitless.
\STATE This signal is modeled as a single gear with $2N$ teeth.
\STATE The corresponding diameter is $T_{1{\rm{/}}N} / 0.275$ unitless.
\ELSIF {Implement a clock signal of $M/N$ duty cycle.}
\STATE Three gears are totally required.
\STATE \({G_1}\) denotes $1/2$, and \({G_2}\) with $2N$ teeth denotes $1/N$.
\STATE \({G_3}\) is concentric with \({G_2}\), painted by two colors.
\STATE One is pink for $M$ sections, the other is grey for $N\!\!-\!\!M$ ones.
\ENDIF
\end{algorithmic}
\end{algorithm}

\begin{figure*}[htbp]
\centering
\includegraphics[width=0.7\linewidth]{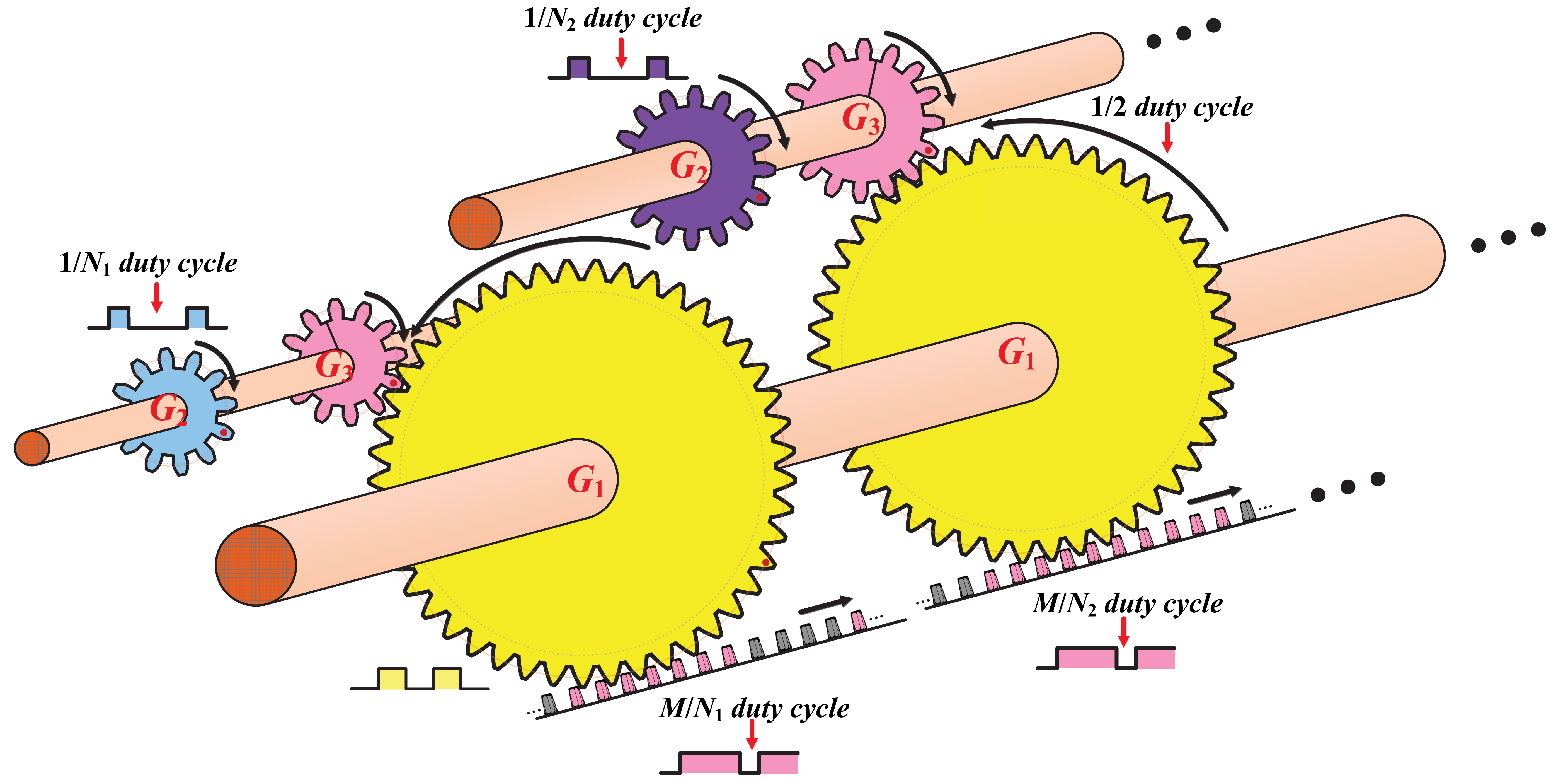}
\caption{Gear model of a clock tree.}\label{fig5}
\end{figure*}

\subsection{For a Clock Tree}
A clock tree synthesized with CRNs indicates a lot of clock signals with various duty cycles. The operating mechanism works as what Fig. \ref{fig5} shows. Various gears of \({G_3}\) mesh with the corresponding \({G_1}\) of $ 1/2 $ duty cycle. All the concentric gears of \({G_1}\) share the same diameters as well as the clock time period. Thus, for each sub-system, composed of \({G_1}\), \({G_2}\), and \({G_3}\), different clock signals of $1/N_i$ produce different $M/N_i$, where $i$ is a positive integer index .

\begin{figure}[htbp]
\centering
\includegraphics[width=0.97\linewidth]{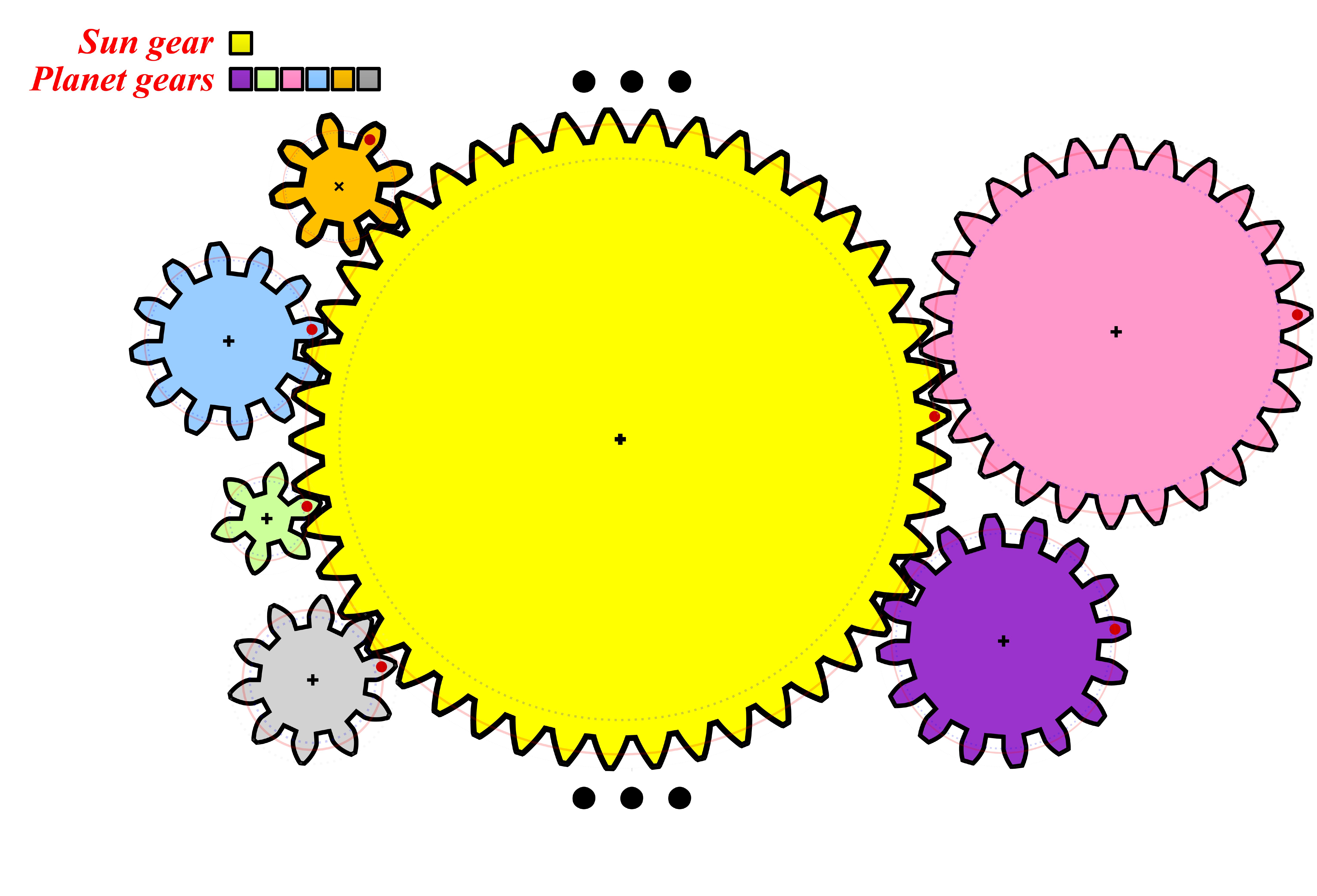}
\caption{Side view of the gear model for a clock tree. The biggest gear colored yellow is the sun gear, and others are the planet ones. All the planet gears rotate around the sun gear, transmitting the clock information.}\label{sq3}
\end{figure}

More interestingly, the side view of Fig. \ref{fig5} looks like various gears that rotate around a single gear, extremely similar to those different planets move around the sun. Fig. \ref{sq3} displays this mentioned side view. In the center of the whole system is a yellow gear \textbf{(the sun)}, namely \({G_1}\), representing for $1/2$ duty cycle. Around \({G_1}\) are a lot of painted \({G_3}\) \textbf{(the planets)}, representing for different $1/N$ ($N > 2$) clock signals. All \({G_2}\) are concentric with their corresponding \({G_3}\) with the same gear size, thereby sharing the same time period.

\begin{Rem}\label{rem:7}
Based on the analysis in Section \ref{subsec:3a}, we know the diameter of \({G_1}\) is usually larger than \({G_2}\), hence there are a large number of planets could be engaged with \({G_1}\). Most clock signals could be produced through this mechanism.
\end{Rem}

\begin{figure}[htbp]
\centering
\subfigure[Gear model of $2/3$ duty cycle.]{
\label{fig4:a}
\includegraphics[width=0.9\linewidth]{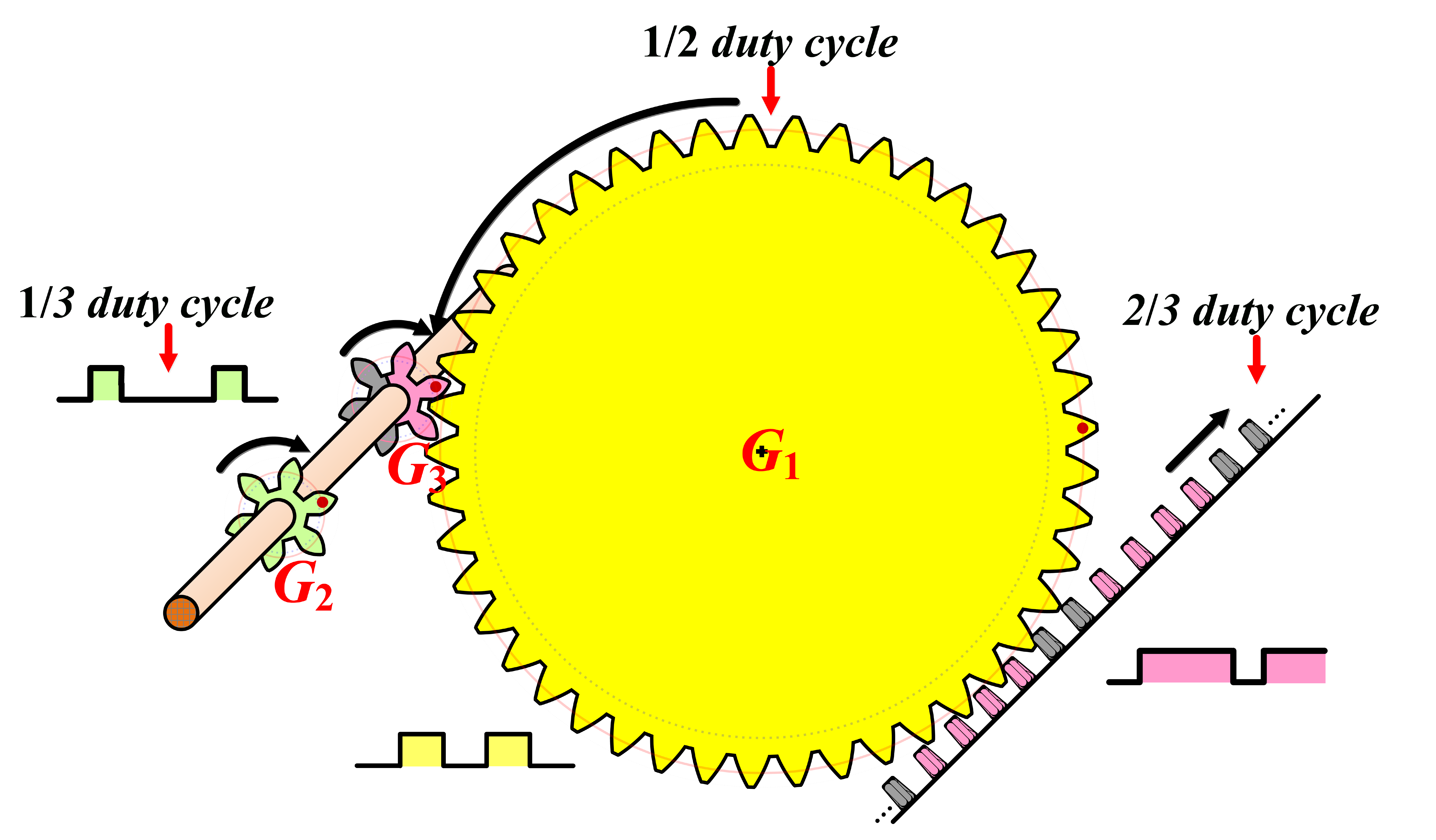}}\\
\subfigure[Simulation result for $2/3$ duty cycle.]{
\label{fig4:b}
\includegraphics[width=0.75\linewidth]{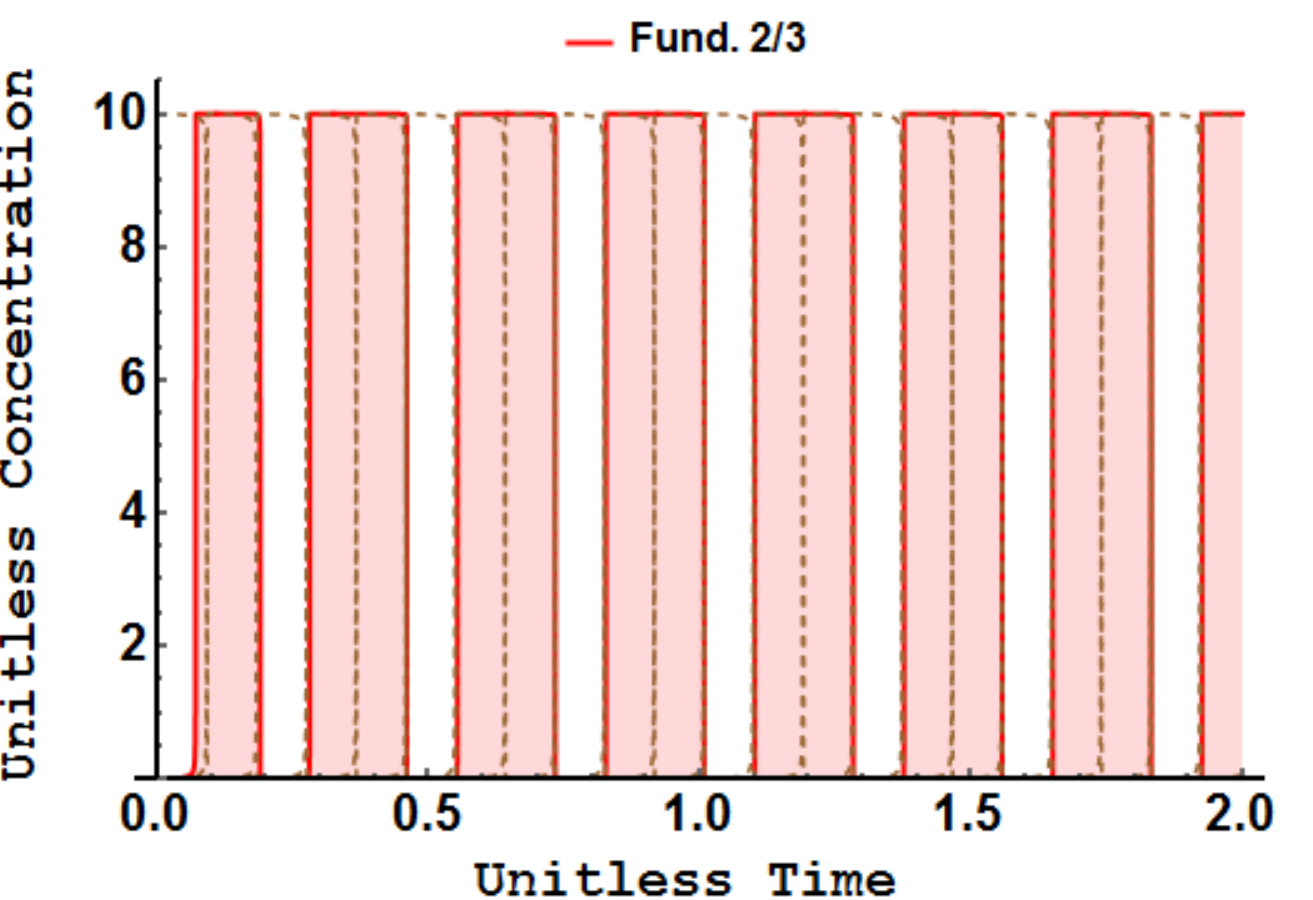}}
\caption{Implementation and simulation result for $2/3$ duty cycle.}\label{fig4}
\end{figure}

\textbf{\textit{Example.}} Take a $2/3$ duty cycle clock signal as an example. After the implementation of a rock and pinion as Fig. \ref{fig4:a}, the time has been recorded in the rock just as the simulation results shown in Fig. \ref{fig4:b}.

Actually, a skew is found in Fig. \ref{fig4:b} at the very beginning of simulation time. This skew can be attributed to the first touch or mesh between \({G_1}\) and \({G_3}\). After that, they mesh well with each other and nearly no skew would be produced.

\section{Discussion and Conclusion}\label{sec:6}
To our best knowledge, we are the first to borrow ideas from gears to synthesize a tunable clock signal with CRNs. In this paper, inspired by a gear system, we could better understand the operating mechanism of a clock signal with different duty cycles in CRNs. In our previous work \cite{ge2015formal}, nearly no quantitative description of the results was offered, however. This paper semi-quantitatively reproduces all the simulation data with a set of physically reasonable parameters. On a chassis of our gear model, a good understanding of a clock tree in CRNs could be obtained from the standpoint of signal generation, information transport as well as the unwanted clock skew. The proposed clock synthesis methodology based on gear systems can helps designers conveniently derive the required clock signals. Experimental works with DNA strand displacement reactions will be shown in our future work.

\section*{Acknowledgment}
First, we would like to thank Dr. David Soloveichik for his kind help. We also would like to thank Editor-in-Chief, the associate editor, and the reviewers for their time and efforts to review this paper.

\appendices
\section{Concrete Chemical Reactions}\label{sec:app}
\subsection{Reactions for RGB Oscillators}
Concrete chemical reactions are shown in Table \ref{t1} with totally $12$ reactions. Attention should be paid to the different rate constants of reactions with various functionalities.

\begin{table}[ht]
\centering
\caption{Reactions for the RGB oscillators}
\begin{tabular}{P{17pt}||P{25pt}|P{43pt}|P{49pt}|P{49pt}}
\Xhline{1.0pt}
\textbf{Clock} & \textbf{Absence} & \textbf{Phase signal} & \textbf{Threshold} & \textbf{Main power}\\
\hline
\textbf{Rate} &  {$k=1$} & $k=100$ & $k=1$ & $k=100$\\
\hline
\hline
~&{\cellcolor{mygray}\(\phi \to \textcolor[rgb]{1,0,0}{r}\)}&{\cellcolor{mygray}\(\textcolor[rgb]{1,0,0}{R}+\textcolor[rgb]{1,0,0}{r} \to \textcolor[rgb]{1,0,0}{R}\)}&{\cellcolor{mygray}\(\textcolor[rgb]{1,0,0}{R}+\textcolor[rgb]{0,0,1}{b} \to \textcolor[rgb]{0,1,0}{G}+\textcolor[rgb]{0,0,1}{b}\)}&{\cellcolor{mygray}\(\textcolor[rgb]{1,0,0}{R}+2\textcolor[rgb]{0,1,0}{G} \to 3\textcolor[rgb]{0,1,0}{G}\)}\\
RGB&\(\phi \to \textcolor[rgb]{0,1,0}{g}\)&\(\textcolor[rgb]{0,1,0}{G}+\textcolor[rgb]{0,1,0}{g} \to \textcolor[rgb]{0,1,0}{G}\)&\(\textcolor[rgb]{0,1,0}{G}+\textcolor[rgb]{1,0,0}{r} \to \textcolor[rgb]{0,0,1}{B}+\textcolor[rgb]{1,0,0}{r}\)&\(\textcolor[rgb]{0,1,0}{G}+2\textcolor[rgb]{0,0,1}{B} \to 3\textcolor[rgb]{0,0,1}{B}\)\\
~&{\cellcolor{mygray}\(\phi \to \textcolor[rgb]{0,0,1}{b}\)}&{\cellcolor{mygray}\(\textcolor[rgb]{0,0,1}{B}+\textcolor[rgb]{0,0,1}{b} \to \textcolor[rgb]{0,0,1}{B}\)}&{\cellcolor{mygray}\(\textcolor[rgb]{0,0,1}{B}+\textcolor[rgb]{0,1,0}{g} \to \textcolor[rgb]{1,0,0}{R}+\textcolor[rgb]{0,1,0}{g}\)}&{\cellcolor{mygray}\(\textcolor[rgb]{0,0,1}{B}+2\textcolor[rgb]{1,0,0}{R} \to 3\textcolor[rgb]{1,0,0}{R}\)}\\
\Xhline{1.0pt}
\end{tabular}\label{t1}
\end{table}

\subsection{Reactions for 1/2 Duty Cycle Clock Signal}
Concrete chemical reactions are shown in Table \ref{t2} with wholly $12$ chemical reactions. Note the reactions of instructor production and annihilation, especially the rate constant of main power reactions.

\begin{table}[ht]
\centering
\caption{Reactions for $1/2$ duty cycle clock}
\begin{tabular}{c|c||c}
\Xhline{1.0pt}
\textbf{State} & \textbf{Rate} & \(cl{k_1}\) \\
\hline
\rowcolor{mygray}
\textbf{Absence} &  $k=1$ & \(\phi  \to i{n_1}\) \\
\textbf{Phase signal}  &  $k=1$ & \(cl{k_1} + i{n_1} \to cl{k_1}\) \\
\rowcolor{mygray}
\textbf{Instructors}  &  $k=1$ & \(cl{k_1} + i{n_0} \to cl{k_1} + {t_{10}}\)  \\
\textbf{Threshold } &  $k=1$ & \({t_{10} }+ cl{k_1} \to cl{k_0} + {t_{10}}\) \\
\rowcolor{mygray}
\textbf{Main power}  &  $k=1000$ &\({t_{10}} + 2cl{k_0} + cl{k_1} \to 3cl{k_0} + {t_{10}}\)  \\
\textbf{Annihilation}  &  $k=1$ & \({t_{10}} \to \phi \) \\
\Xhline{1.0pt}
\textbf{State} & \textbf{Rate} & \(cl{k_0}\) \\
\hline
\rowcolor{mygray}
\textbf{Absence}  &  $k=1$ & \(\phi  \to i{n_0}\) \\
\textbf{Phase signal}   &  $k=1$ & \(cl{k_0} + i{n_0} \to cl{k_0}\)\\
\rowcolor{mygray}
\textbf{Instructors}  &  $k=1$ & \(cl{k_0} + i{n_1} \to cl{k_0} + {t_{01}}\) \\
\textbf{Threshold } &  $k=1$  & \({t_{01}} + cl{k_0} \to cl{k_1} + {t_{01}}\) \\
\rowcolor{mygray}
\textbf{Main power}  &  $k=1000$ & \({t_{01}} + 2cl{k_1} + cl{k_0} \to 3cl{k_1} + {t_{01}}\) \\
\textbf{Annihilation}  &  $k=1$ & \({t_{01}} \to \phi \) \\
\Xhline{1.0pt}
\end{tabular}
\label{t2}
\vspace*{4pt}
\end{table}
\subsection{Reactions for Frequency Division 1/15}
 Concrete chemical reactions are shown in Eq. \ref{eq1:1} with wholly $72$ chemical reactions, rate constant of which remains the same with reactions in Tables \ref{t1} and \ref{t2}.

\begin{subequations}\label{eq1:1}
\begin{equation}\label{subeq1:1:1}
    \left\{
    \begin{aligned}
        \varphi  &\to r,\\
        \varphi  &\to {r_1},\\
        \varphi  &\to {r_2},\\
        \varphi  &\to {r_3},\\
        \varphi  &\to {r_4};
    \end{aligned}
    \right.
    \left\{
    \begin{aligned}
        \varphi  &\to {r_5},\\
        \varphi  &\to {r_6},\\
        \varphi  &\to {r_7},\\
        \varphi  &\to {r_8},\\
        \varphi  &\to {r_9};
    \end{aligned}
    \right.
    \left\{
    \begin{aligned}
        \varphi  &\to {r_{10}},\\
        \varphi  &\to {r_{11}},\\
        \varphi  &\to {r_{12}},\\
        \varphi  &\to {r_{13}},\\
        \varphi  &\to {r_{14}};
    \end{aligned}
    \right.
    \end{equation}
\begin{equation}\label{subeq1:1:2}
\resizebox{.88\hsize}{!}{$
    \left\{
    \begin{aligned}
      R + r &\to R,\\
     {R_1} + {r_1} &\to {R_1},\\
     {R_2} + {r_2} &\to {R_2},\\
     {R_3} + {r_3} &\to {R_3},\\
     {R_4} + {r_4} &\to {R_4};
    \end{aligned}
    \right.
    \left\{
    \begin{aligned}
     {R_5} + {r_5} &\to {R_5},\\
     {R_6} + {r_6} &\to {R_6},\\
     {R_7} + {r_7} &\to {R_7},\\
     {R_8} + {r_8} &\to {R_8},\\
     {R_9} + {r_9} &\to {R_9};
    \end{aligned}
    \right.
    \left\{
    \begin{aligned}
     {R_{10}} + {r_{10}} &\to {R_{10}},\\
     {R_{11}} + {r_{11}} &\to {R_{11}},\\
     {R_{12}} + {r_{12}} &\to {R_{12}},\\
     {R_{13}} + {r_{13}} &\to {R_{13}},\\
     {R_{14}} + {r_{14}} &\to {R_{14}};
    \end{aligned}
    \right.
    $}
    \end{equation}
    \begin{equation}\label{subeq1:1:3}
    \left\{
    \begin{aligned}
     R + {r_{14}}  &\to {R_1} + {r_{14}},\\
     {R_1} + r     &\to {R_2} +  r,\\
     {R_2} + {r_1} &\to {R_3} + {r_1},\\
     {R_3} + {r_2} &\to {R_4} + {r_2},\\
     {R_4} + {r_3} &\to {R_5} + {r_3};
    \end{aligned}
    \right.
    \left\{
    \begin{aligned}
     {R_5} + {r_4} &\to {R_6} + {r_4},\\
     {R_6} + {r_5} &\to {R_7} + {r_5},\\
     {R_7} + {r_6} &\to {R_8} + {r_6},\\
     {R_8} + {r_7} &\to {R_9} + {r_7},\\
     {R_9} + {r_8} &\to {R_{10}} + {r_8};
    \end{aligned}
    \right.
     \end{equation}
   \begin{equation}\label{subeq1:1:4}
    \left\{
    \begin{aligned}
     {R_{10}} + {r_9} &\to {R_{10}}   + {r_9},\\
     {R_{11}} + {r_{10}} &\to {R_{11}}+ {r_{10}}  ,\\
     {R_{12}} + {r_{11}} &\to {R_{12}}+ {r_{11}},\\
     {R_{13}} + {r_{12}} &\to {R_{13}}+ {r_{12}},\\
     {R_{14}} + {r_{13}} &\to {R_{14}}+ {r_{13}};
    \end{aligned}
    \right.
    \left\{
    \begin{aligned}
     R + 2{R_1}     &\to 3{R_1},\\
     {R_1} + 2{R_2} &\to 3{R_2} ,\\
     {R_2} + 2{R_3} &\to 3{R_3} ,\\
     {R_3} + 2{R_4} &\to 3{R_4} ,\\
     {R_4} + 2{R_5} &\to 3{R_5} ;
    \end{aligned}
    \right.
    \end{equation}
    \begin{equation}
    \left\{
    \begin{aligned}
     {R_5} + {r_6} &\to 3{R_6} ,\\
     {R_6} + {r_7} &\to 3{R_7},\\
     {R_7} + {r_8} &\to 3{R_8},\\
     {R_8} + {r_9} &\to 3{R_9} ,\\
     {R_9} + {r_{10}} &\to 3{R_{10}};
    \end{aligned}
    \right.
    \left\{
    \begin{aligned}
     {R_{10}} + 2{R_{11}} &\to 3{R_{11}},\\
     {R_{11}} + 2{R_{12}} &\to 3{R_{12}},\\
     {R_{12}} + 2{R_{13}} &\to 3{R_{13}},\\
     {R_{13}} + 2{R_{14}} &\to 3{R_{14}},\\
     {R_{14}} + R &\to 3R;
    \end{aligned}
    \right.
    \end{equation}
     \begin{equation}\label{subeq1:1:6}
    \left\{
    \begin{aligned}
    \phi  &\to i{n_0},\\
    cl{k_0} + i{n_0} &\to cl{k_0},\\
    cl{k_0} + i{n_1} &\to cl{k_0} + {t_{01}},\\
R+  {t_{10}} + cl{k_1} &\to  R + cl{k_0} + {t_{10}},\\
R+    {t_{10}} + 2cl{k_0} + cl{k_1} &\to  R+ 3cl{k_0} + {t_{10}};\\
    {t_{10}} &\to \phi;
    \end{aligned}
    \right.
    \end{equation}
    \begin{equation}\label{subeq1:1:7}
    \left\{
    \begin{aligned}
     \phi  &\to i{n_1},\\
    cl{k_1} + i{n_1} &\to cl{k_1},\\
    cl{k_1} + i{n_0} &\to cl{k_1} + {t_{10}},\\
{R_{14}}+   {t_{01}} + cl{k_0} &\to {R_{14}} + cl{k_1} + {t_{01}},\\
{R_{14}}+   {t_{01}} + 2cl{k_1} + cl{k_0} &\to  {R_{14}} + 3cl{k_1} + {t_{01}},\\
   {t_{01}} &\to \phi.
    \end{aligned}
    \right.
    \end{equation}
     \end{subequations}


\ifCLASSOPTIONcaptionsoff
  \newpage
\fi
\footnotesize
\bibliographystyle{IEEEtran}
\flushend
\bibliography{IEEEabrv,mybib}

\begin{thebibliography}{10}
\providecommand{\url}[1]{#1}
\csname url@samestyle\endcsname
\providecommand{\newblock}{\relax}
\providecommand{\bibinfo}[2]{#2}
\providecommand{\BIBentrySTDinterwordspacing}{\spaceskip=0pt\relax}
\providecommand{\BIBentryALTinterwordstretchfactor}{4}
\providecommand{\BIBentryALTinterwordspacing}{\spaceskip=\fontdimen2\font plus
\BIBentryALTinterwordstretchfactor\fontdimen3\font minus
  \fontdimen4\font\relax}
\providecommand{\BIBforeignlanguage}[2]{{%
\expandafter\ifx\csname l@#1\endcsname\relax
\typeout{** WARNING: IEEEtran.bst: No hyphenation pattern has been}%
\typeout{** loaded for the language `#1'. Using the pattern for}%
\typeout{** the default language instead.}%
\else
\language=\csname l@#1\endcsname
\fi
#2}}
\providecommand{\BIBdecl}{\relax}
\BIBdecl

\bibitem{synbio}
\BIBentryALTinterwordspacing
Wikipedia. [Online]. Available:
  \url{https://en.wikipedia.org/wiki/Synthetic_biology}
\BIBentrySTDinterwordspacing

\bibitem{kosuri2014large}
S.~Kosuri and G.~M. Church, ``Large-scale de novo \protect{DNA} synthesis:
  Technologies and applications,'' \emph{Nature Methods}, vol.~11, no.~5, pp.
  499--507, 2014.

\bibitem{blight2000efficient}
K.~J. Blight, A.~A. Kolykhalov, and C.~M. Rice, ``Efficient initiation of
  \protect{HCV RNA} replication in cell culture,'' \emph{Science}, vol. 290,
  no. 5498, pp. 1972--1974, 2000.

\bibitem{gibson2008complete}
D.~G. Gibson, G.~A. Benders, C.~Andrews-Pfannkoch, E.~A. Denisova,
  H.~Baden-Tillson, J.~Zaveri, T.~B. Stockwell, A.~Brownley, D.~W. Thomas,
  M.~A. Algire \emph{et~al.}, ``Complete chemical synthesis, assembly, and
  cloning of a \protect{Mycoplasma} genitalium genome,'' \emph{Science}, vol.
  319, no. 5867, pp. 1215--1220, 2008.

\bibitem{luo2000synthetic}
D.~Luo and W.~M. Saltzman, ``Synthetic \protect{DNA} delivery systems,''
  \emph{Nature Biotechnology}, vol.~18, no.~1, pp. 33--37, 2000.

\bibitem{broglia1998folding}
R.~Broglia, G.~Tiana, S.~Pasquali, H.~Roman, and E.~Vigezzi, ``Folding and
  aggregation of designed proteins,'' \emph{Proceedings of the National Academy
  of Sciences (PNAS)}, vol.~95, no.~22, pp. 12\,930--12\,933, 1998.

\bibitem{broglia1999stability}
R.~Broglia, G.~Tiana, H.~Roman, E.~Vigezzi, and E.~Shakhnovich, ``Stability of
  designed proteins against mutations,'' \emph{Physical Review Letters},
  vol.~82, no.~23, p. 4727, 1999.

\bibitem{church2012next}
G.~M. Church, Y.~Gao, and S.~Kosuri, ``Next-generation digital information
  storage in \protect{DNA},'' \emph{Science}, vol. 337, no. 6102, pp.
  1628--1628, 2012.

\bibitem{eigen1966chemical}
M.~Eigen and L.~Maeyer, ``Chemical means of information storage and readout in
  biological systems,'' \emph{Naturwissenschaften}, vol.~53, no.~3, pp. 50--57,
  1966.

\bibitem{black1987biochemistry}
I.~B. Black, J.~E. Adler, C.~F. Dreyfus, W.~F. Friedman, E.~F. LaGamma, and
  A.~H. Roach, ``Biochemistry of information storage in the nervous system,''
  \emph{Science}, vol. 236, no. 4806, pp. 1263--1268, 1987.

\bibitem{turner1987biosensors}
A.~Turner, I.~Karube, and G.~S. Wilson, ``Biosensors: Fundamentals and
  applications,'' \emph{Oxford University Press}, 1987.

\bibitem{sassolas2008dna}
A.~Sassolas, B.~D. Leca-Bouvier, and L.~J. Blum, ``\protect{DNA} biosensors and
  microarrays,'' \emph{Chemical reviews}, vol. 108, no.~1, pp. 109--139, 2008.

\bibitem{turner2000biosensors}
A.~P. Turner, ``Biosensors--sense and sensitivity,'' \emph{Science}, vol. 290,
  no. 5495, pp. 1315--1317, 2000.

\bibitem{ray1993evolutionary}
T.~S. Ray, ``An evolutionary approach to synthetic biology: Zen and the art of
  creating life,'' \emph{Artificial Life}, vol.~1, no. 1\_2, pp. 179--209,
  1993.

\bibitem{langton1997artificial}
C.~G. Langton, \emph{Artificial life: An overview}.\hskip 1em plus 0.5em minus
  0.4em\relax MIT Press, 1997.

\bibitem{malyshev2014semi}
D.~A. Malyshev, K.~Dhami, T.~Lavergne, T.~Chen, N.~Dai, J.~M. Foster, I.~R.
  Corr{\^e}a, and F.~E. Romesberg, ``A semi-synthetic organism with an expanded
  genetic alphabet,'' \emph{Nature}, vol. 509, no. 7500, pp. 385--388, 2014.

\bibitem{franco2011timing}
E.~Franco, E.~Friedrichs, J.~Kim, R.~Jungmann, R.~Murray, E.~Winfree, and F.~C.
  Simmel, ``Timing molecular motion and production with a synthetic
  transcriptional clock,'' \emph{Proceedings of the National Academy of
  Sciences (PNAS)}, vol. 108, no.~40, pp. E784--E793, 2011.

\bibitem{kim2011synthetic}
J.~Kim and E.~Winfree, ``Synthetic in vitro transcriptional oscillators,''
  \emph{Molecular systems biology}, vol.~7, no.~1, p. 465, 2011.

\bibitem{cardelli2009artificial}
L.~Cardelli, \emph{Artificial biochemistry}.\hskip 1em plus 0.5em minus
  0.4em\relax Springer, 2009.

\bibitem{ge2015formal}
L.~Ge, C.~Zhang, Z.~Zhong, and X.~You, ``A formal design methodology for
  synthesizing a clock signal with an arbitrary duty cycle of \protect{M/N},''
  in \emph{Proc. IEEE Workshop on Signal Processing Systems (SiPS)}, 2015, pp.
  1--6.

\bibitem{litak2005dynamics}
G.~Litak and M.~I. Friswell, ``Dynamics of a gear system with faults in meshing
  stiffness,'' \emph{Nonlinear Dynamics}, vol.~41, no.~4, pp. 415--421, 2005.

\bibitem{walha2009nonlinear}
L.~Walha, T.~Fakhfakh, and M.~Haddar, ``Nonlinear dynamics of a two-stage gear
  system with mesh stiffness fluctuation, bearing flexibility and backlash,''
  \emph{Mechanism and Machine Theory}, vol.~44, no.~5, pp. 1058--1069, 2009.

\bibitem{gear}
\BIBentryALTinterwordspacing
D.~J. Dunn. [Online]. Available:
  \url{http://www.freestudy.co.uk/dynamics/gears.pdf}
\BIBentrySTDinterwordspacing

\bibitem{soloveichik2010dna}
D.~Soloveichik, G.~Seelig, and E.~Winfree, ``\protect{DNA} as a universal
  substrate for chemical kinetics,'' \emph{Proceedings of the National Academy
  of Sciences (PNAS)}, vol. 107, no.~12, pp. 5393--5398, 2010.

\bibitem{zhang2010dynamic}
D.~Y. Zhang, ``Dynamic \protect{DNA} strand displacement circuits,'' Ph.D.
  dissertation, California Institute of Technology, 2010.

\bibitem{zhang2011dynamic}
D.~Y. Zhang and G.~Seelig, ``Dynamic \protect{DNA} nanotechnology using
  strand-displacement reactions,'' \emph{Nature Chemistry}, vol.~3, no.~2, pp.
  103--113, 2011.

\bibitem{dnadis}
\BIBentryALTinterwordspacing
M.~Research. [Online]. Available:
  \url{http://research.microsoft.com/apps/video/default.aspx?id=228977}
\BIBentrySTDinterwordspacing

\bibitem{cardelli2013two}
L.~Cardelli, ``Two-domain \protect{DNA} strand displacement,''
  \emph{Mathematical Structures in Computer Science}, vol.~23, no.~02, pp.
  247--271, 2013.

\bibitem{jiang2010synthesis}
H.~Jiang, A.~Kharam, M.~Riedel, and K.~K. Parhi, ``A synthesis flow for digital
  signal processing with biomolecular reactions,'' in \emph{Proc. IEEE
  International Conference on Computer-Aided Design (ICCAD)}, 2010, pp.
  417--424.

\bibitem{jiang2011synchronous}
H.~Jiang, M.~Riedel, and K.~Parhi, ``Synchronous sequential computation with
  molecular reactions,'' in \emph{Proc. IEEE/ACM Design Automation Conference
  (DAC)}, 2011, pp. 836--841.

\bibitem{llg}
L.~Ge, Z.~Zhong, D.~Wen, X.~You, and C.~Zhang, ``A formal combinational logic
  synthesis with chemical reaction networks,'' \emph{IEEE Transactions on
  Molecular, Biological and Multi-Scale Communications}, vol.~3, no.~1, pp.
  33--47, 2017.

\bibitem{horn1972general}
F.~Horn and R.~Jackson, ``General mass action kinetics,'' \emph{Archive for
  Rational Mechanics and Analysis}, vol.~47, no.~2, pp. 81--116, 1972.

\bibitem{mass}
\BIBentryALTinterwordspacing
Wikipedia. [Online]. Available:
  \url{https://en.wikipedia.org/wiki/Law_of_mass_action}
\BIBentrySTDinterwordspacing

\bibitem{phillips2009programming}
A.~Phillips and L.~Cardelli, ``A programming language for composable
  \protect{DNA} circuits,'' \emph{Journal of the Royal Society Interface},
  vol.~6, no. Suppl 4, pp. S419--S436, 2009.

\bibitem{NUPACK}
\BIBentryALTinterwordspacing
Caltech. [Online]. Available: \url{http://www.nupack.org/}
\BIBentrySTDinterwordspacing

\bibitem{cardelli2008processes}
L.~Cardelli, ``From processes to \protect{ODEs} by chemistry,'' in \emph{Proc.
  International Conference on Theoretical Computer Science (TCS)}, 2008, pp.
  261--281.

\bibitem{kharam2011binary}
A.~Kharam, H.~Jiang, M.~Riedel, and K.~Parhi, ``Binary counting with chemical
  reactions.'' in \emph{Proc. Pacific Symposium on Biocomputing}, 2011, pp.
  302--313.

\bibitem{pl}
\BIBentryALTinterwordspacing
stackoverflow. [Online]. Available:
  \url{http://mathematica.stackexchange.com/questions/19859/plot-extract-data-to-a-file}
\BIBentrySTDinterwordspacing

\end{thebibliography}

\end{document}